\documentclass[]{aa}

\def\Msol{\hbox{M$_\odot$}}
\def\Msun{\hbox{M$_\odot$}}
\def\Zsun{\hbox{Z$_\odot$}}
\def\Zsol{\hbox{Z$_\odot$}}

\def\kms{\hbox{km$\,$s$^{-1}$}}
\def\cmt{\hbox{cm$^{-3}$}}
\def\Apix{\hbox{\AA$\,$pix$^{-1}$}}
\def\one{\,{\sc i}}             
\def\two{\,{\sc ii}}
\def\three{\,{\sc iii}}
\def\four{\,{\sc iv}}

\def\fsec{\hbox{$.\!\!^{\rm s}$}}

\def\micron{\textmu m}

\newcommand{\hei}{He\,{\sc i}}

\newcommand{\foi}{[O\,{\sc i}]}
\newcommand{\foii}{[O\,{\sc ii}]}
\newcommand{\fsii}{[S\,{\sc ii}]}
\newcommand{\fsiii}{[S\,{\sc iii}]}
\newcommand{\foiii}{[O\,{\sc iii}]}
\newcommand{\fnii}{[N\,{\sc ii}]}
\newcommand{\ffeiii}{[Fe\,{\sc iii}]}

\newcommand{\ha}{H$\alpha$}
\newcommand{\hb}{H$\beta$}

\newcommand{\elt}{$T_{\rm e}$}
\newcommand{\eld}{$N_{\rm e}$}

\usepackage[varg]{txfonts}
\usepackage{amsmath}
\usepackage{amssymb}
\usepackage{mathrsfs}
\usepackage{color}
\usepackage{soul} 
\usepackage{textgreek}
\usepackage{graphicx}
\usepackage{overpic}
\usepackage{booktabs,threeparttable}
\newcommand{\mtnote}[1]{\textsuperscript{\TPTtagStyle{#1}}}   
\usepackage{fancyvrb}
\usepackage{chngpage}

\begin{document}
   \title{Piecing together the puzzle of NGC~5253: abundances, kinematics and WR stars\thanks{Based on observations with the Gemini-South telescope under programme GS-2008A-Q-25.}}


   \author{M.\ S.\ Westmoquette\inst{1}
          \and
          B.\ James\inst{2,3}
          \and
          A.\ Monreal-Ibero\inst{4}
          \and
          J.\ R.\ Walsh\inst{1}
          }

   \institute{European Southern Observatory, Karl-Schwarzschild-Str. 2, 85748 Garching bei M\"{u}nchen, Germany\\
              \email{mwestmoq@eso.org}
         \and
             Space Telescope Science Institute, 3700 San Martin Drive, Baltimore, MD 21218, USA\\
         \and
             Institute of Astronomy, University of Cambridge, Madingley Road, Cambridge, CB3 0HA, UK\\
         \and
             Instituto de Astrof\'{i}sica de Andaluc\'{i}a (CSIC), C/ Camino Bajo de Hu\'{e}tor, 50, 18008 Granada, Spain
             }

   \date{Received ; accepted 10 December 2012}

\VerbatimFootnotes

\abstract{
We present Gemini-S/GMOS-IFU optical spectroscopy of four regions near the centre of the nearby (3.8~Mpc) dwarf starburst galaxy NGC~5253. This galaxy is famous for hosting a radio supernebula containing two deeply embedded massive super star clusters, surrounded by a region of enhanced nitrogen abundance that has been linked to the presence of WR stars. We detected 11 distinct sources of red WR bump (C\four) emission over a 20$''$ ($\sim$350~pc) area, each consistent with the presence of $\sim$1 WCE-type star. WC stars are not found coincident with the supernebula, although WN stars have previously been detected here. We performed a multi-component decomposition of the H$\alpha$ line across all four fields and mapped the kinematics of the narrow and broad (FWHM=100--250~\kms) components. These maps paint a picture of localised gas flows, as part of multiple overlapping bubbles and filaments driven by the star clusters throughout the starburst. We confirm the presence of a strong H$\alpha$ velocity gradient over $\sim$$4\farcs5$ ($\sim$80~pc) coincident with the region of N/O enhancement, and high gas density known from previous study, and interpret this as an accelerating ionized gas outflow from the supernebula clusters. We measure the ionized gas abundances in a number of regions in the outer IFU positions and combine these with measurements from the literature to assess the radial abundance distribution. We find that the O/H and N/H profiles are consistent with being flat. Only the central 50~pc exhibits the well-known N/O enhancement, and we propose that the unusually high densities/pressures in the supernebula region have acted to impede the escape of metal-enriched hot winds from the star clusters and allow them to mix with the cooler phases, thus allowing these freshly processed chemicals to be seen in the optical.}


\titlerunning{Piecing together the puzzle of NGC 5253}
\maketitle

\section{Introduction}

Local starbursts are the present-day manifestations of the intense star formation thought to have been instrumental in building galaxies in the early Universe. Starbursts occur in dense, gas-rich environments where the structure and kinematics of the ionized medium reflect the intense energy input from massive stars via stellar winds and supernova explosions. These winds can disrupt the surrounding gas, drive material out from the galaxy, and trigger (or quench) further star formation. 

\object{NGC 5253} is an excellent nearby \citep[distance of 3.8 Mpc;][]{sakai04} example of a starburst. This Blue Compact Dwarf (BCD) has ongoing active star formation at a rate of $\sim$0.2~\Msol~yr$^{-1}$ \citep{calzetti07, lopez-sanchez12}, and contains a large number of compact young ($\sim$1--12~Myr) star clusters \citep{gorjian96, harris04, cresci05, vanzi06} located within the central $\sim$250 pc starburst zone. Optical studies have been able to identify only very few clusters older than ~20 Myr \citep{tremonti01, harris04}, supporting the extreme youth of this starburst. Recently three potentially massive ($\gtrsim$$10^5$~\Msun) and old (1--2~Gyr) star clusters have been found in the outskirts of the galaxy underlining the fact that the current starburst is just the most recent star forming episode \citep{harbeck12}. \citet{lopez-sanchez12} concluded that the starburst was most likely triggered by the infall of an H\one\ cloud \citep{meier02}, although they could not rule out a contribution from ram pressure stripping as the galaxy moves through the dense intergalactic medium of the Centaurus A group \citep{karachentsev07}.

H$\alpha$ images show a centrally concentrated region extending $\sim$30$''$ in the north-south direction, surrounded by a chaotic ionized structure with radial filaments, bubbles and loops \citep{marlowe95, martin98, calzetti99}. Dense clouds within these structures produce rapidly varying levels of extinction. The morphology of the X-ray emission that traces the high-temperature gas follows the H$\alpha$ distribution closely, and is consistent with a high energy-density outflow similar to that found in the dwarf starburst galaxy NGC 1569 \citep{strickland99, summers04}.

Within the region of brightest H$\alpha$ emission are embedded two very massive (combined mass $\sim$1--2$\times$$10^6$~\Msol), young \citep[age $\lesssim$4~Myr, but possibly $\lesssim$6~Myr;][]{martin-hernandez05}, compact super star clusters (SSCs) separated by $\sim$$0\farcs4$ \citep{vanzi04, alonsoherrero04}. These clusters exist behind $A_{V}$=7--17 magnitudes of extinction \citep{vanzi04, alonsoherrero04}, and are connected with (and possibly co-spatial with) a very strong mid-IR \citep{gorjian01} and radio emission source \citep{turner00, turner04}. Hereafter we will refer to this central giant H\two\ region as the ``supernebula'' \citep[following][]{turner00}.

NGC~5253 is classified as a low-metallicity system \citep[$\sim$0.2--0.3~\Zsun;][]{kobulnicky99} with a fairly uniform oxygen distribution. Famously, however, a factor of 2--3 nitrogen enhancement in the central regions has been found in a number of targeted studies \citep{walsh89, kobulnicky97a, lopez-sanchez07}. Evidence for the presence of nitrogen-rich WN- and carbon-rich WC-type Wolf-Rayet stars \citep{schaerer97} in the same N-enriched regions led to the suggestion that these WR stars could be the cause of this enhancement \citep{walsh89, kobulnicky97a}. Indeed, in general, WR galaxies (those with large WR populations) as observed with  SDSS samples exhibit elevated N/O ratios relative to non-WR galaxies \citep{brinchmann08}.

\citet{monreal10a, monreal12} presented the results of the first spatially-resolved study of the ionized gas kinematics and abundances in the central regions of NGC~5253 with VLT/FLAMES integral field unit (IFU) observations. They found that, while the abundances of O, Ne and Ar remain constant within $\lesssim$0.1~dex over the mapped area, there are in fact two regions of elevated N/O: one strong enhancement associated with the supernebula, and one weaker one associated with two older ($\sim$10~Myr) clusters towards the south-west. The fact that signatures of WR stars are not found at the location of this second enhancement led to the conclusion that there must be a more complex relationship between the presence of WR stars and N/O excess than previously thought. Interestingly, \citet{james13} find high N/O ratios in Haro~11 in a region with an integrated age just older than the WR phase ($\sim$7~Myr), but normal N/O ratios in two regions with large WR star populations, implying that the mixing time-scales are fairly long.

In this paper we present optical IFU observations of four regions within the NGC~5253 starburst, including the supernebula region and two additional fields 150--250~pc away. We begin by presenting the distribution of carbon-rich Wolf-Rayet (WR) stars (WC-type stars; via the $\sim$5800~\AA\ red WR bump) in all four regions (Section~\ref{sect:wr}), since this has not been examined previously. We then present a detailed study of the kinematical properties of the ionized gas in all four regions (Section~\ref{sect:kinematics}) and the nitrogren and oxygen distribution in the two outer regions (Section~\ref{sect:abund}), and in comparison to measurements from the literature. A detailed study of the He\one\ abundance distribution within the central regions will be presented in a forthcoming paper (Monreal-Ibero et al.\ in prep.). Finally we discuss the state of the ISM across the whole starburst region, including the correspondence between the kinematics, chemical composition and presence of WR stars in relation to the enigmatic supernebula.



\section{Observations}

In 2008 queue-mode observations using the Gemini-South Multi-Object Spectrograph (GMOS) Integral Field Unit \citep[IFU;][]{allington02} in one-slit mode were obtained of four regions near the centre of NGC 5253 (programme ID: GS-2008A-Q-25, PI: M.\ Westmoquette), in 0.5--0.7~arcsec seeing\footnote{Position 1 was observed in exceptional seeing of $0\farcs46$.} ($\sim$9--12~pc at the distance of NGC~5253) and photometric conditions. A nearby bright star was used to provide guiding and tip-tilt corrections using the GMOS on-instrument wave front sensor \mbox{(OIWFS)}.

The one-slit mode IFU gives a field-of-view (FoV) of $5\times 3.5$~arcsec ($\sim$$90\times 60$~pc) sampled by 500 contiguous hexagonal lenslets of $0\farcs2$ diameter. We took a set of dithered exposures at each position, offsetting by integer spaxel values (i.e.\ $0\farcs2$) and used the R831 grating to give a spectral coverage of 4750--6850~\AA\ at a dispersion of 0.34~\Apix. An additional block of 250 lenslets (covering $5\times 1.7$~arcsec), offset by 1~arcmin from the object field, provides a dedicated sky view. Positions 1, 2 and 4 were tilted by 5$^{\circ}$ in order to avoid light from the galaxy from contaminating the sky fibres. Table~\ref{tbl:gmos_obs} lists details of the observations for each position.

The GMOS spectrograph is fed by optical fibres from the IFU which reformats the arrangement of the spectra for imaging by the detector. In order to remove the fibre-to-fibre throughput and detector pixel-to-pixel sensitivity differences, and enable the wavelength and flux calibration of the data, a number of bias frames, flat-fields, twilight flats, arc calibration frames, and observations of the photometric standard star Hiltner 600, were obtained as part of the standard baseline calibration plan.

\begin{figure*}
\centering
\includegraphics[width=0.75\textwidth]{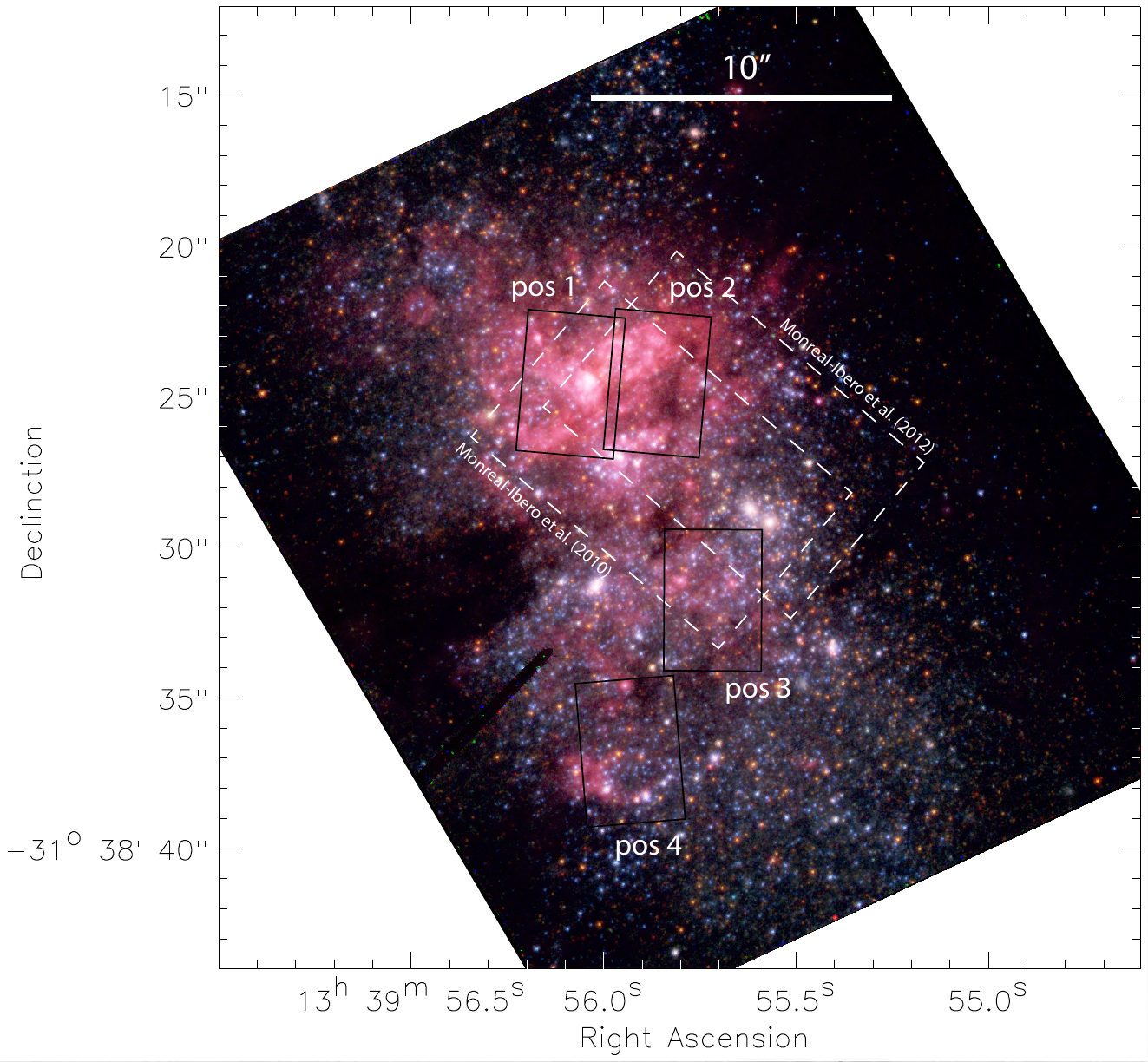}
\caption{\textit{HST}/ACS-HRC colour composite (F330W, F435W, F550M, F814W, F658N) image of NGC 5253 showing the location of the four GMOS-IFU positions. The dashed boxes indicate the location of the VLT/FLAMES-ARGUS IFU pointings from \citet{monreal10a} and \citet{monreal12} for comparison.}
\label{fig:finder}
\end{figure*}

The IFU positions were chosen to cover selected areas of the disturbed ionized interstellar medium in the inner regions of NGC~5253, covering the nebulosity surrounding the central SSCs \citep{turner00, alonsoherrero04} and a further two regions to the south containing a number of ionized clumps. In Fig.~\ref{fig:finder}, we show the position of the IFU fields on an \textit{HST}/ACS-HRC colour composite image (PID 10609, P.I.\ Vacca).

\begin{table*}
\centering
\caption {Gemini/GMOS-IFU observation log}
\label{tbl:gmos_obs}
\begin{tabular}{c c r @{\hspace{0.2cm}} l c r @{ $\times$ } l c}
\hline
Pos. & Date & \multicolumn{2}{c}{Coordinates} & PA & \multicolumn{2}{c}{Exp.\ Time} & Seeing \\
No. & & \multicolumn{2}{c}{(J2000)} & ($^{\circ}$) & \multicolumn{2}{c}{(s)} \\
\hline 
1 & 15/2/08 & $13^{\rm h}\,39^{\rm m}\,56\fsec10$ & $-31^{\circ}\,38'\,24\farcs8$ & 355 & 1500 & 4 & $\sim$$0\farcs5$ \\
2 & 1--2/3/08 & $13^{\rm h}\,39^{\rm m}\,55\fsec87$ & $-31^{\circ}\,38'\,24\farcs7$ & 355 & 1500 & 4  & $\sim$$0\farcs5$ \\
3 & 1/3/08, 9/7/08 & $13^{\rm h}\,39^{\rm m}\,55\fsec72$ & $-31^{\circ}\,38'\,31\farcs8$ & 0 & 1500 & 6  & $\sim$$0\farcs6$ \\
4 & 3/3/08, 9/7/08 & $13^{\rm h}\,39^{\rm m}\,55\fsec93$ & $-31^{\circ}\,38'\,36\farcs9$ & 5 & 1500 & 6  & $0\farcs5$--$0\farcs7$ \\
\hline\\
\end{tabular}
\end{table*}

\subsection{Reduction} \label{sect:reduction}
The field-to-slit mapping for the GMOS IFU reformats the layout of the fibres of the sky to one long row containing blocks of 100--150 object fibres interspersed by blocks of 50 sky fibres. In this way, all 750 spectra (500 object + 250 sky) can be recorded on the detector simultaneously.

Basic data reduction was performed following the standard Gemini reduction pipeline (implemented in \textsc{iraf}\footnote{The Image Reduction and Analysis Facility ({\sc iraf}) is distributed by the National Optical Astronomy Observatories which is operated by the Association of Universities for Research in Astronomy, Inc. under cooperative agreement with the National Science Foundation.}). Briefly, a trace of the position of each spectrum on the CCD was first produced from the flat field. Then throughput correction functions and wavelength calibration solutions were created and applied to the science and standard star data, and the averaged sky spectrum (computed from the separate sky field) was subtracted, before the individual spectra were extracted using the flat-field trace. The spectra were then flux calibrated from observations of the standard star. Although all our observations were taken under photometric conditions, that the standard star was not observed on the same night as our science observations means that the accuracy of the spectrophotometric flux calibration cannot be absolute. To test the accuracy of the \emph{relative} flux calibration across the field, we re-reduced one science frame for IFU position 1 without sky subtraction (but with the standard fibre-to-fibre throughput correction based on the twilight flat), and measured the flux of the prominent [O\one]$\lambda$5577 sky line across the whole field. Very little variation was found with no systematic patterns; the mean flux was measured to be $1.17\times10^{-16}$~erg~s$^{-1}$~cm$^{-2}$ with a standard deviation of $8.80\times10^{-18}$~erg~s$^{-1}$~cm$^{-2}$ (i.e.\ a $\sim$7.5\% variation). This is at a level sufficient for our science goals.

The data were converted into standard `cube' format using \textsc{gfcube} with a spatial (over)sampling rate of $0\farcs1$, and corrected for the effects of differential atmospheric correction \citep[using an \textsc{iraf}-based code written by J.\ R.\ Walsh based on an algorithm described in][]{walsh90}, Finally, the individual dithered exposures for each position were combined using a custom \textsc{pyraf} script written by J.\ Turner (Gemini).

In order to determine an accurate measurement of the instrumental contribution to the line broadening, we fitted single Gaussians to isolated arc lines on a wavelength calibrated arc exposure, and found the resolution to be $1.2$~\AA\ at H$\alpha$ (55~\kms) and 1.3~\AA\ at H$\beta$ (80~\kms). The fit residuals shows that the instrumental line profile is extremely close to a Gaussian.

\begin{figure*}
\centering
\includegraphics[width=\textwidth]{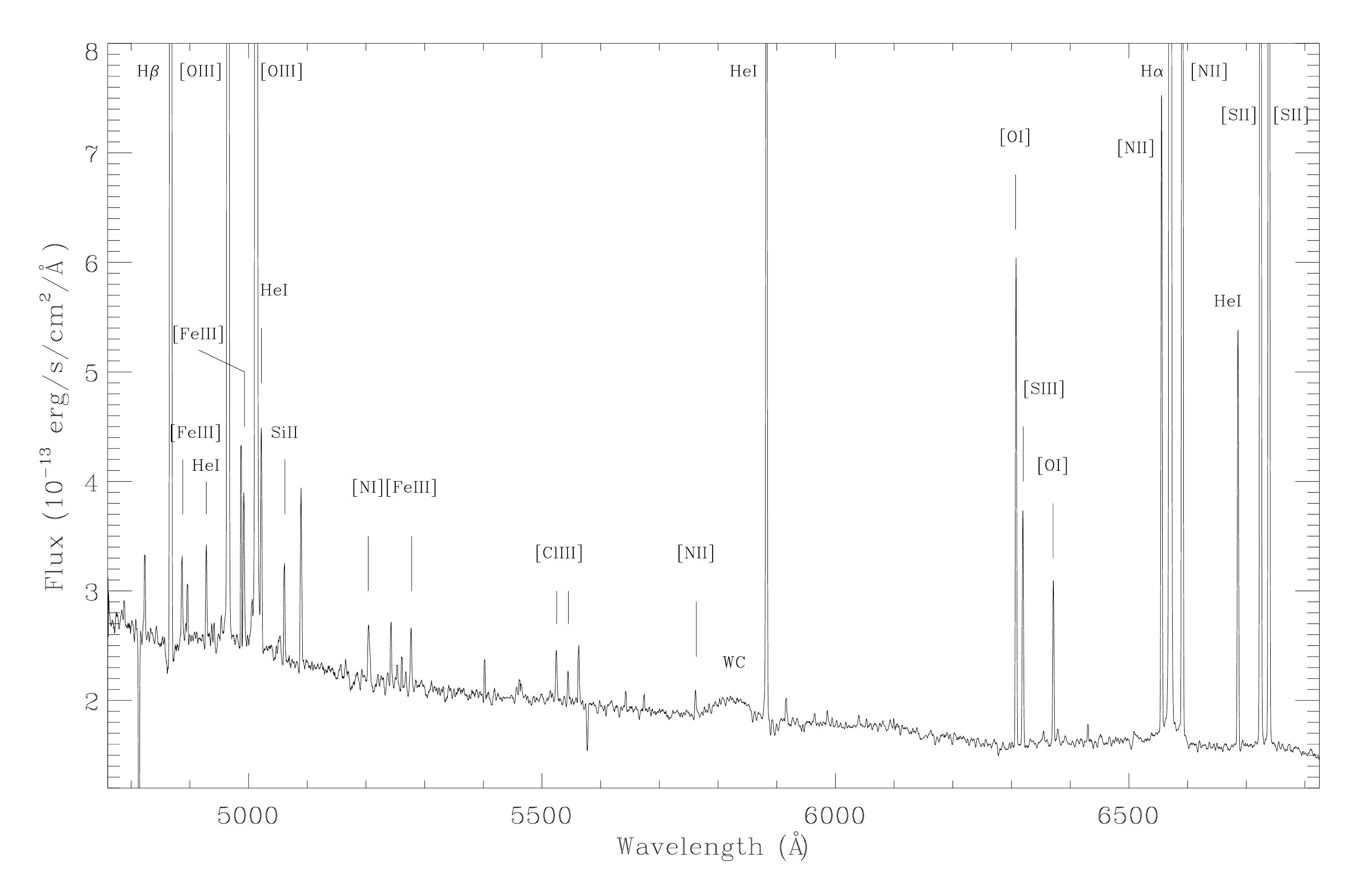}
\caption{Example spectrum summed over 13 spaxels covering a bright ionized knot in IFU position 3 (defined as region 1; see Section~\ref{sect:abund}). All identified emission lines are labelled, including the broad C\four\ $\lambda$5808 feature originating in the atmospheres of WC-type Wolf-Rayet stars.}
\label{fig:spectrum}
\end{figure*}

Fig.~\ref{fig:spectrum} shows an example reduced and flux calibrated spectrum extracted from a 13-spaxel region covering a bright ionized knot in IFU position 3 (which we define as region 1 in Section~\ref{sect:abund}). A plethora of emission lines are detectable from neutral to doubly ionized species, and in this spectrum the red Wolf-Rayet bump is also clearly seen. Line identifications were made using the lists compiled by \citet{tsamis03} and \citet{lopez-sanchez07}.

\section{The carbon-rich Wolf-Rayet star population} \label{sect:wr}

Because of the link posited in previous studies between WR stars and nitrogen enhancements, it is of particular interest to understand the population and distribution of WR stars within NGC~5253. Here we discuss our results on the WC star population, together with other WR star detections from the literature, which together provide a view of the WC and WN star population within the central $\sim$350~pc of the galaxy.

As exemplified in Fig.~\ref{fig:spectrum}, we detect strong WR-star signatures in some spectra in the form of the red WR bump. This feature is a blend of the broad C\three\ $\lambda$5698 and C\four\ $\lambda$5808 emission lines originating in the atmospheres of carbon-rich WC-type WR stars. At low metallicities such as in NGC~5253, early-type WC stars (WCE-types) are expected to be the dominant sub-type \citep{crowther08}.

\begin{figure*}
\begin{minipage}{\textwidth}
\centering
\includegraphics[width=0.5\textwidth]{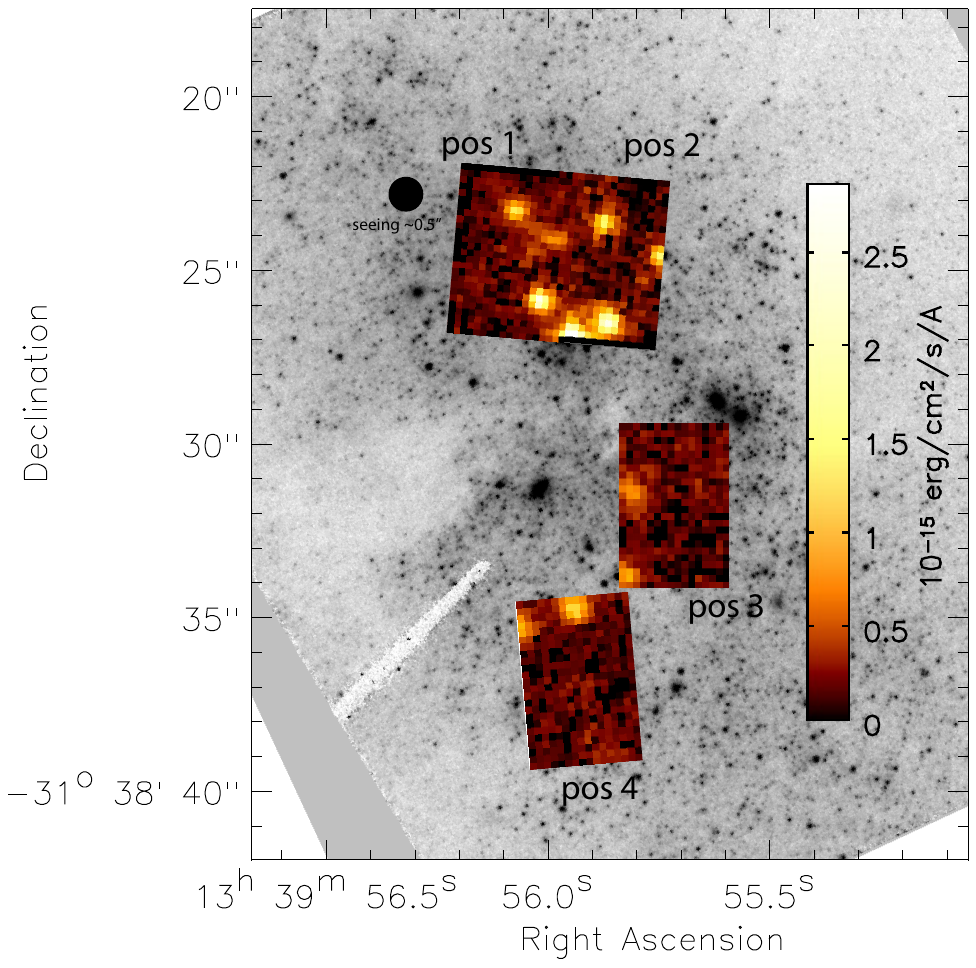}
\includegraphics[width=0.4\textwidth]{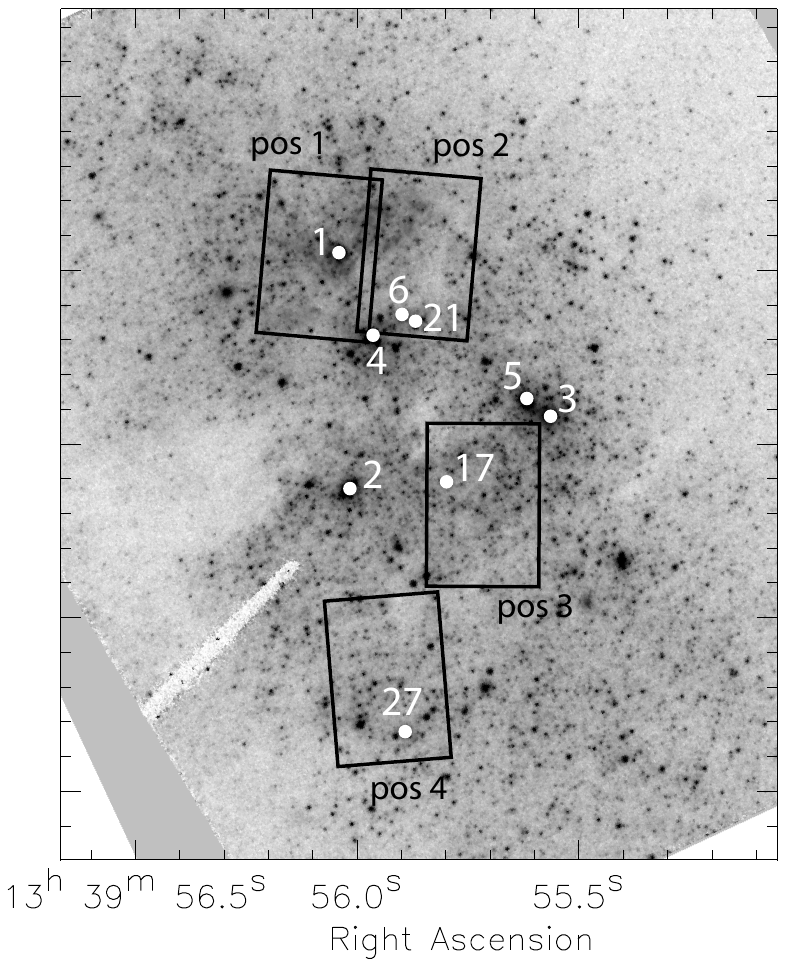}
\end{minipage}
\caption{\textit{Left:} Integrated maps of the C\four\ $\lambda$5808 and C\three\ $\lambda$5698 WC-star emission feature (the red WR bump) for each position on the same colour scale \textbf{(as shown)}, overlaid on an \textit{HST}/ACS-HRC F550M image (ID: 10609, PI: Vacca). The emission flux in each knot \textbf{(consistent with a point source the size of the seeing disk)} is consistent with $\sim$1 WCE-type star. The HRC F550M image is reproduced on the right with a number of the clusters identified by \citet{harris04}, and discussed in the text, labelled.}
\label{fig:finder_WR}
\end{figure*}

We mapped out the strength and location of the red WR bump in our four IFU fields. First we measured the continuum level in two windows either side of the WR feature and subtracted the average. We then collapsed the datacubes over the wavelength range 5765--5875~\AA\ to produce integrated maps of the red WR bump for all four positions. These are shown in Fig.~\ref{fig:finder_WR}. The size of each WR emission region is consistent with being a point source (i.e.\ the size of the seeing disk, $\sim$0.5--0.7~arcsecs). To estimate the number of WR stars in each source, we used the relation given by \citep{lopez-sanchez10} for the C\four\,$\lambda$5808 luminosity of a WCE star for a given metallicity (their equation 8). At the metallicity and distance of NGC~5253, this relation predicts a WCE star luminosity of $2.3\times 10^{36}$~erg~s$^{-1}$; given the uncertainties in both our measured fluxes and the WCE star luminosity relation, each source is consistent with $\sim$1--few WCE stars (see Table~\ref{tab:abundances}).

For a typical cluster mass of $5\times 10^3$~\Msol\ \citep{harris04} and age of 4~Myr (Table~\ref{tbl:age_WR}), the number of WR stars predicted by Starburst99 \citep{leitherer99} is $\sim$2. Firstly, this is in good agreement with our finding of $\sim$1 WR star per source. Secondly, this very low number means that the presence of WR stars in any given cluster is dominated by stochasticity. Not every cluster of the right age and mass will contain a WR star (i.e.\ the IMF is not fully sampled in clusters of this mass). Indeed, simple Monte Carlo sampling of a Chabrier IMF over a total mass of $5\times 10^3$~\Msol, shows that only $\sim$1--10 stars $>$30~\Msol\ are expected.

We do not detect a red WR bump at the location of the supernebula (the peak of emission in the optical; cluster H04-1\footnote{Hereafter we refer to the clusters identified in the study of \citet{harris04} with the notation H04-X, where X is their cluster ID. We note that their catalogue is by no means complete. Their study was based on WFPC2 data; more recent ACS-HRC imaging resolves very many more clusters (e.g.\ Fig.~\ref{fig:finder_WR}).}), but find a number of WC stars within 3$''$ including one coincident with cluster H04-21, and a concentration at the UV emission peak (cluster H04-4). We also detect four more WR emission sources much further from the supernebula in GMOS positions 3 and 4.

\citet{schaerer97} detected W-R features (from both WN- and WC-type stars) at both near the peak of emission in the optical and the UV, equivalent to $\sim$10 WC stars and a few 10s of WN stars in each of the two regions. Due to a pointing error during their observations they did not in fact cover the optical and UV flux peaks (clusters H04-1 and 4), but positioned their slit $\sim$1\farcs$5$ to the west (D. Schaerer, priv. comm. 2007).
\citet{sidoli10}, however, did obtain optical long-slit spectroscopy of the optical and UV peaks with VLT/UVES, and detected 5 WN and 1 WC star in a 1$''$$\times$3$''$ region covering the supernebula (cluster H04-1), and 10 WN and 2 WC stars at the UV peak (cluster H04-4).
\citet{lopez-sanchez07} found $\sim$1 WN star each in their regions A, B (a few arcsecs west and north-west of the supernebula), and $\sim$10 WN stars in region C (equivalent to cluster H04-4). 
Accounting for the different distances and WR subtype luminosities adopted by \citet{lopez-sanchez07}, these values are in reasonable agreement with those found by \citet{sidoli10} in the corresponding areas.
\citet{monreal10a} identified seven distinct regions containing blue-bump emission from WN-type stars in their IFU data covering the nuclear regions. The location of their detections are consistent with those found by the above authors. 

\begin{figure}
\centering
\includegraphics[width=0.5\textwidth]{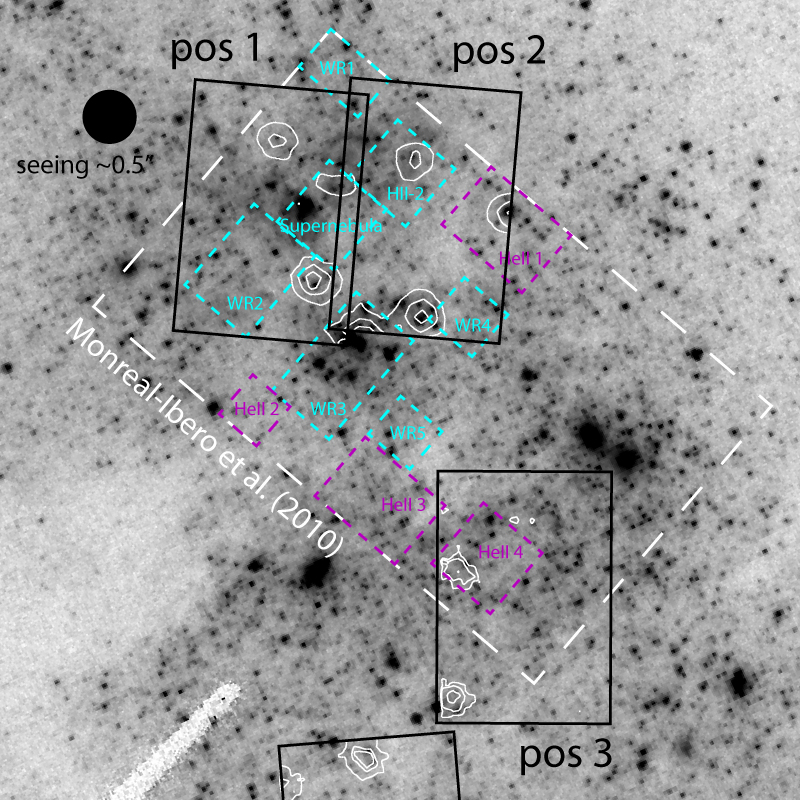}
\caption{Zoom in to the central region of NGC~5253 showing all the available IFU data on the location of the WR star population. Within GMOS IFU positions 1, 2 and 3, the red WR bump flux is represented with white contours. The size of the seeing disk for positions 1 and 2 ($0\farcs5$) is shown in the top left. The large white dashed rectangle shows the VLT/FLAMES-ARGUS IFU position of \citet{monreal10a}, and the labelled cyan and magenta rectangles show the regions in which evidence for a blue WR bump and He\two\ emission was found, respectively.}
\label{fig:finder_WRpos12}
\end{figure}

Fig.~\ref{fig:finder_WRpos12} shows a zoom-in of the central regions of NGC~5253 summarising the IFU data on the location of the WC star population from this study and the WN stars from \citet{monreal10a}. In only two locations are there coincident detections of WC and WN stars (regions WR3 and H\two~2 of \citealt{monreal10a}), although the spatial overlap of our datasets is far from complete. Of the four regions in which nebular He\two\ (i.e.\ narrow emission, as opposed to the very broad emission originating in the WR stellar atmosphere) is detected by \citet{monreal10a}, two are covered by our GMOS data, and both contain a WC star (regions He\two~1 and He\two~4 of \citealt{monreal10a}; see Fig.~\ref{fig:finder_WRpos12}). Only certain classes of WR stars, and possibly certain O stars, are predicted to be hot enough to ionize nebular He\two\ \citep{kudritzki02, crowther07}, whose ionization potential is $h\nu = 54.4$~eV, and so far only one WC star has been associated with a He\two\ nebula \citep[WC star MC45 in nebula BCLMP38b in M33;][]{kehrig11}. Our spectroscopy shows that both our He\two-WC regions have higher than average [O\three]/H$\beta$ line ratios, consistent with the presence of a hard radiation field. Neither, however, show convincing evidence for the presence of shocks in the traditional BPT diagrams \citep[][not shown]{baldwin81}.

Almost all previous studies of the WR population in NGC~5253 have focussed on the central regions. However, \citet{sidoli10} examined a region $\sim$5$''$ to the north of the supernebula, and found evidence for 1 WN star, and \citet{lopez-sanchez07} found evidence for a further WN star in their region D (coincident with cluster H04-2, $\sim$10$''$ south of the supernebula and $\sim$5$''$ to the east of GMOS IFU position 3). We detect red-bump WR emission consistent with the presence of single WC stars in 4 locations in GMOS positions 3 and 4. In position 3, one of these WC stars is coincident with H04-17, a known cluster of mass 2--3$\times$10$^{3}$~\Msol\ and age $<$4~Myr \citep{harris04}. The remaining 3 sources are not coincident with any previously identified clusters. Table~\ref{tbl:age_WR} summarises the detections of WN and WC stars in the regions defined in both this study and that of \citet{monreal10a}, where we have ordered them by decreasing EW(H$\beta$) (or increasing age using the predictions of Starburst99).

For all stars that pass through a WR phase, it is thought that the WN stage precedes the WC stage \citep{crowther07}. Thus, one would expect WN stars to be present predominantly in the younger regions, and WC stars in the older regions. Disregarding regions 2 and 6 (where we do not detect a red WR bump, and that were not covered by the observations of \citealt{monreal10a}), there does appear to be a transition at $\sim$~3.4--4.4 Myr, below which there is a  higher probability of finding WN stars, and above which there is a higher probability of finding WC star. Clear exceptions to this are regions H\two\ 2 (very young but exhibits evidence for both WR types) and WR5 (which, according to its H$\beta$ flux, is older but has WN star signatures). In neither of these locations did \citet{harris04} identify a cluster, but as we have previously mentioned, their catalogue is far from complete. Furthermore, \citet{monreal10a} were only able to determine the location of any WR emission to a precision of $\sim$$1\farcs5$, which is not sufficient to be sure which source (if identified) is responsible. Projection effects also add to the difficulties of disentangling which source is associated with which.

In summary, we detect 11 distinct sources of C\four\,$\lambda$5808 emission throughout the NGC~5253 starburst, each consistent with the presence of $\sim$1 WCE type star. Some are associated with known young clusters catalogued by \citet[][although this study is by no means complete]{harris04}, and two are coincident with the location of WN stars detected by \citet{monreal10a}. The finding of WR stars spread out over 20$''$ ($\sim$350~pc) attests to the large area over which the starburst has occurred.

\begin{table*}
\centering
\caption{H$\beta$ equivalent widths (EW), ages and presence of WR stars in regions from both this study (bold numbers; see Table~\ref{tab:abundances} for details of the measurements) and that of \citet{monreal10a}.}
\label{tbl:age_WR}
\begin{scriptsize}
\begin{threeparttable}
\begin{tabular}{l ccccccccccccccc}
\hline
Region & \textbf{SN}\tnote{a} & H\two\ 2 & WR2 & WR1 & WR3 & WR4 && \textbf{5} & \textbf{6} & \textbf{1}, He\two~4 & WR5 & \textbf{2} & \textbf{4} & \textbf{3} & He\two\ 1 \\
Other literature & H04-1, &&&& H04-4, && H04-21 &&& H04-17 \\
\hspace{0.2cm}names\tnote{b} & LS-B &&&& LS-C &&&&&  \\
EW(H$\beta$) (\AA) & 250--350 & 320 & 180 & 130 & 130 & 130 & 110 & 105 & 100 & 95 & 85 & 60 & 34 & 14 & 9 \\
Age (Myr)\tnote{c} & 2.3--2.7 & 2.4 & 3.2 & 3.4 & 3.4 & 3.4 & 4.4 & 4.5 & 4.6 & 4.6 & 4.7 & 4.8 & 5.4 & 6.3 & 7.5 \\
WN\tnote{d} & \checkmark & \checkmark & \checkmark & \checkmark & \checkmark & \checkmark & $\times$ & ? & ? & $\times$ & \checkmark & ? & ? & ? &  $\times$ \\
WC\tnote{d} & $\times$ & \checkmark & ? & ? & \checkmark & ? & \checkmark & \checkmark & $\times$ & \checkmark & ? & $\times$ & \checkmark & \checkmark & \checkmark \\
\hline
\end{tabular}
\begin{tablenotes}
\item[a] SN=Supernebula. The range in EWs and ages reflects the variation in values found in the literature.
\item[b] H04=\citealt{harris04}, LS=\citealt{lopez-sanchez07}, MI=\citealt{monreal10a}.
\item[c] Ages calculated from the predictions of Starburst99 \citep{leitherer99}.
\item[d] ? indicates where the relevant WR signature has not been observed in this region.
\end{tablenotes}
\end{threeparttable}
\end{scriptsize}
\end{table*}

\section{Emission line kinematics} \label{sect:kinematics}

Previous long-slit observations have shown that the ionized gas kinematics in the central regions of NGC~5253 are complex, with line profiles revealing multiple components and asymmetric wings \citep[e.g.][]{walsh89, martin95, vanzi06, lopez-sanchez07, sidoli10}. \citet{monreal10a} present a detailed analysis of the emission line kinematics in the central regions as covered by their VLT/FLAMES-ARGUS observations. They found two--three components were needed to fit the observed H$\alpha$ line profile, and attributed them (at least near the central SSC) to cluster winds and multiple expanding shells.

As mentioned above, \citet{sidoli10} presented VLT/UVES long-slit echelle spectroscopy of the optical and UV peaks \citep[knots 1 and 2 from the study of][]{monreal10a}. They also found three Gaussian components were required to satisfactorily fit the emission line profiles: a narrow (FWHM=30--40~\kms) component, a broad (FWHM=100--150~\kms, blueshifted by 10--30~\kms\ with respect to C1) and a third, fainter component intermediate in width and radial velocity. With a spectral resolution of R$\sim$28\,000 ($\sim$11~\kms\ at H$\alpha$), the fact that they still needed a broad component suggests that it is not simply a convolution of many narrower, unresolved components \citep[c.f.][for 30Dor]{melnick99}.

\subsection{Decomposing the line profiles} \label{sect:line_profiles}
We fitted multiple Gaussian profile models to the H$\alpha$ emission line using an \textsc{idl} $\chi^{2}$ fitting package called \textsc{pan} \citep[Peak ANalysis;][]{dimeo}, to recover kinematic information about the ionized gas in the four IFU positions. The high spectral resolution and signal-to-noise (S/N) of our data, together with the fact that the instrumental line profile is so close to a Gaussian (Section~\ref{sect:reduction}), have allowed us to quantify the H$\alpha$ line profile shape to a high degree of accuracy. We can therefore be confident of the identification of faint components and/or those with very small ($\lesssim$10~\kms) velocity separations.

To account for the variable number of Gaussian components in each line profile, we fitted a single, double, and triple Gaussian component initial guess to each. Line fluxes were constrained to be positive and widths to be greater than the instrumental contribution to guard against spurious results. Multi-component fits were run several times with different initial guess configurations (widths and wavelengths) and the fit with the lowest $\chi^{2}$ value was kept. However, we note that the $\chi^2$ minimisation routine employed by \textsc{pan} is very robust with respect to the initial guess parameters. 

To determine how many Gaussian components best fit an observed profile (one, two or three), we used a likelihood ratio test to determine whether a fit of $n$+1 components was more appropriate to an $n$-component fit. This test says that if the ratio of the $\chi^{2}$ statistics for the two fits falls above a certain threshold, then the fits are considered statistically distinguishable, and the one with the lower $\chi^{2}$ can be selected. Here we determine the threshold ratio by visual inspection of a range of spectra and fits. This method is a more generalised form of the formal F-test that we have used in previous work, e.g.\ \citet{westm09a, westm11}. We are aware of work that cautions against the use of the F-test to test for the presence of a line \citep[or additional line component; e.g.][]{protassov02}, but given the absence of a statistically correct alternative that could be applied sensibly to the volume of data presented here, we have chosen to opt for this generalised $\chi^{2}$ ratio test approach.

This test, however, only tells us which of the fits (single, double or triple component) is most appropriate for the corresponding line profile. Experience has taught us that we need to apply a number of additional, physically motivated tests to filter out well-fit but physically improbable results. We specified that the measured FWHM had to be greater than the associated error on the FWHM result (a common symptom of a bad fit), the fluxes of all components should be $>$0, and we rejected any fits where $\chi^{2}_{\rm single}/\chi^{2}_{\rm double}=0$ (another symptom of a bad fit).

\begin{figure*}
\centering
\includegraphics[width=\textwidth]{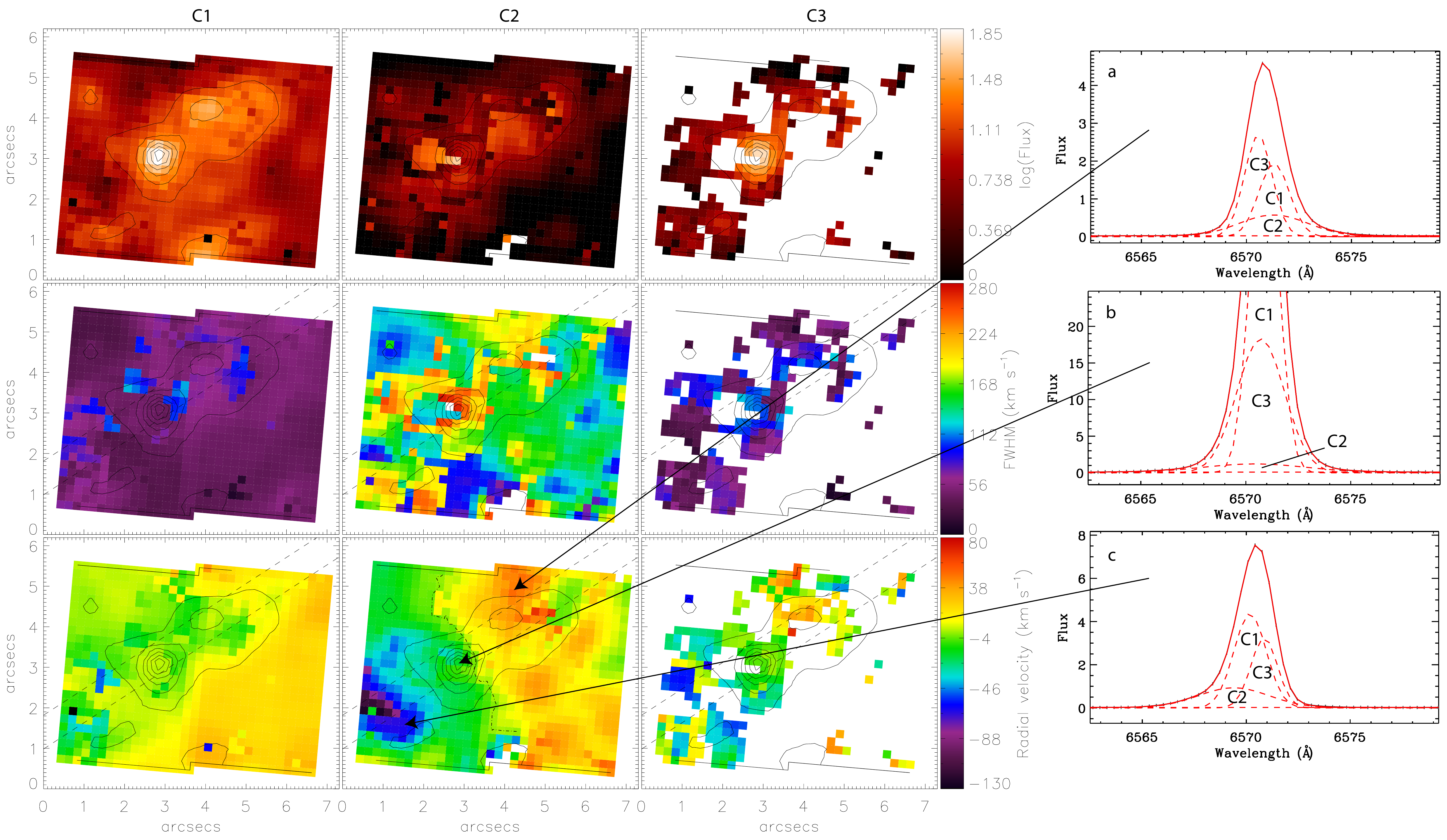}
\caption{Pos 1 and 2 H$\alpha$ flux (top panels), FWHM (middle panels) maps in units of \kms\ (corrected for instrumental broadening), and radial velocity maps (bottom panels) in units of \kms, relative to $v_{\rm sys}$ = $-390$~\kms\ \citep{schwartz04} for the three identified line components. Contours show the summed H$\alpha$ intensity (the peak at coordinates (3,3) is the supernebula). Example H$\alpha$ line profiles and their best multi-component fits are shown for select spaxels as discussed in the text. The parallel dashed lines show the position and width of the pseudoslit we defined to extract the velocity information for plotting in Fig.~\ref{fig:Ha_pv}. The dot-dashed line in the C2 radial velocity map (bottom-centre) shows the zero velocity (i.e.\ $v_{\rm sys}$) contour.}
\label{fig:pos12_Ha_kinem}
\end{figure*}

\begin{figure*}
\centering
\includegraphics[width=\textwidth]{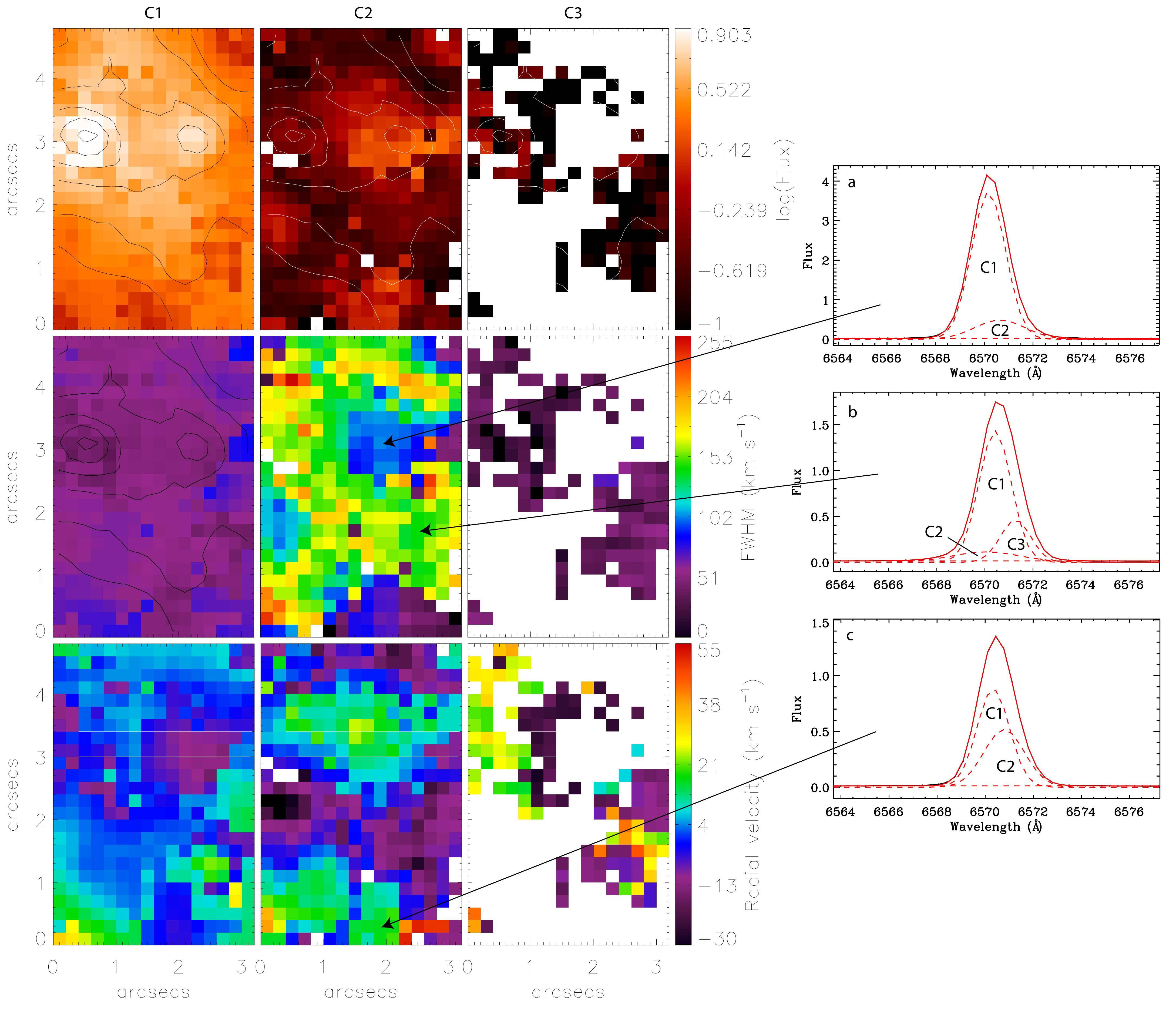}
\caption{Pos 3 H$\alpha$ flux (top panels), FWHM (middle panels) maps in units of \kms\ (corrected for instrumental broadening), and radial velocity maps (bottom panels) in units of \kms, relative to $v_{\rm sys}$ = $-390$~\kms\ \citep{schwartz04} for the three identified line components. Contours show the summed H$\alpha$ intensity. Example H$\alpha$ line profiles and their best multi-component fits are shown for select spaxels as discussed in the text.}
\label{fig:pos3_Ha_kinem}
\end{figure*}

\begin{figure*}
\centering
\includegraphics[width=\textwidth]{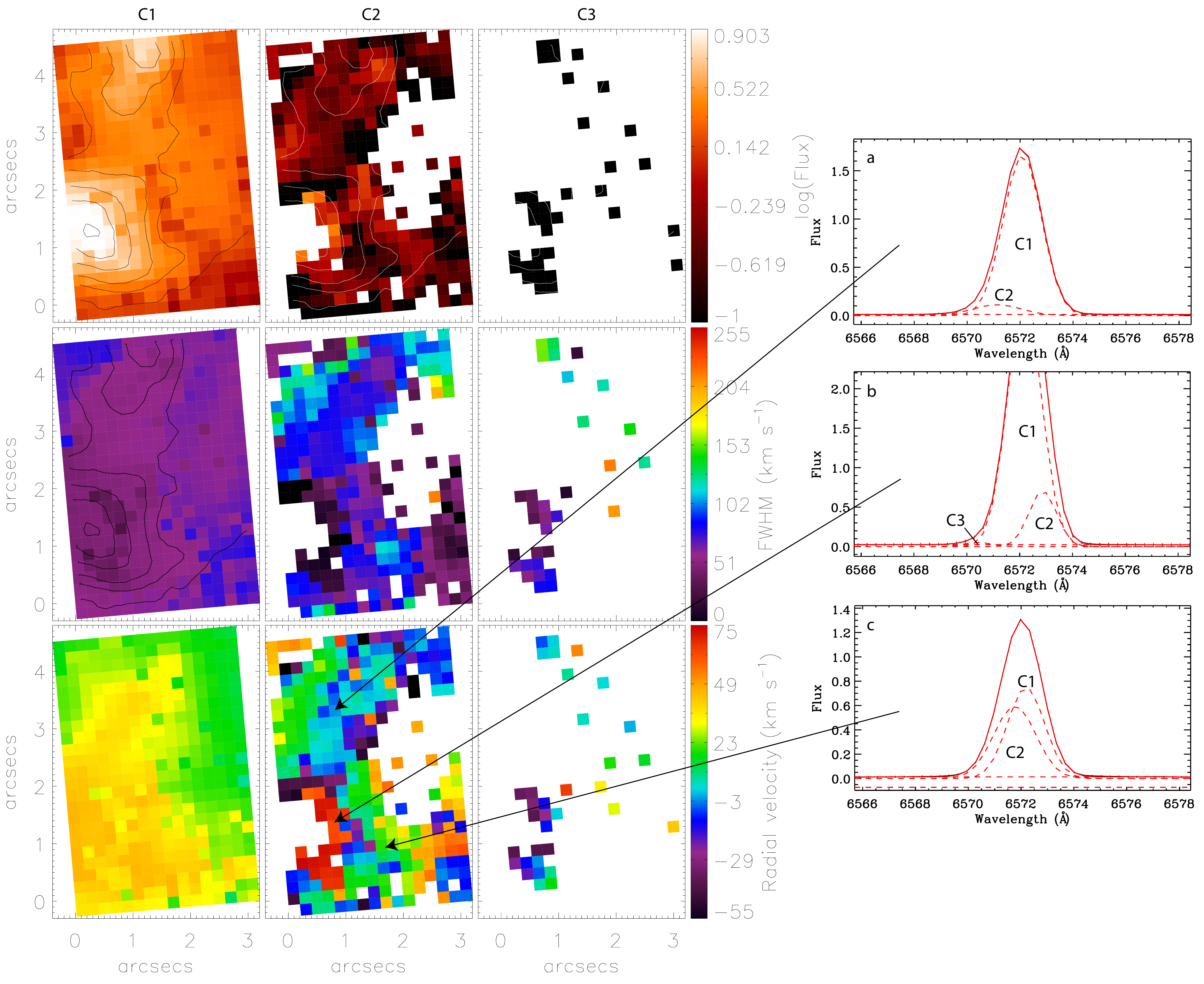}
\caption{Pos 4 H$\alpha$ flux (top panels), FWHM (middle panels) maps in units of \kms\ (corrected for instrumental broadening), and radial velocity maps (bottom panels) in units of \kms, relative to $v_{\rm sys}$ for the three identified line components. Contours show the summed H$\alpha$ intensity.  Example H$\alpha$ line profiles and their best multi-component fits are shown for select spaxels as discussed in the text.}
\label{fig:pos4_Ha_kinem}
\end{figure*}

The final step of the fitting procedure was to assign each fit component to a particular map in such a way as to limit the confusion that might arise during analysis of the results, such as discontinuous spatial regions arising from incorrect component assignments. In positions 1, 2 and 3, we found that the maps were most coherent when we assigned the broadest component to C2, and after that, the brightest as C1 and the faintest to C3. In position 4, we simply assigned the brightest component to C1 (regardless of width), then the faintest component to C3. This approach helps to limit the confusion that might otherwise arise during analysis of the results where discontinuous spatial regions might arise from incorrect component assignments.

Experience has also shown us that the errors that \textsc{pan} quotes on its fit results are an underestimate of the true uncertainties, particularly once uncertainties on our post-fitting tests and filters are taken into account. We can therefore estimate more realistic errors through a visual re-inspection of the profile+fit after knowing which one was selected by our tests, and by taking into account the S/N of the spectrum. We estimate that errors in FWHM range between 0.25--2~\kms\ for C1, 4--20~\kms\ for C2 and 15--25~\kms\ for C3, whereas those for radial velocities range between 0.1--3~\kms\ for C1, 2--5~\kms\ for C2 and 10--30~\kms\ for C3. A 5--10 percent increase in the C1 and C2 uncertainties may occur with the addition of a third component. 

Figs.~\ref{fig:pos12_Ha_kinem}, \ref{fig:pos3_Ha_kinem} and \ref{fig:pos4_Ha_kinem} show the resulting FWHM and radial velocity maps for positions 1, 2, 3 and 4, respectively. In these figures we also include a number of example H$\alpha$ line profiles and best-fitting Gaussian models from certain spaxels. These examples were selected to illustrate the discussion below, but also demonstrate the high quality of the spectra and the accuracy of the line-fitting.

\subsubsection{Positions 1 and 2}
As mentioned above, \citet{monreal10a} present a detailed analysis of the emission line kinematics in the central regions. However, the higher S/N and spatial resolution of our observations permits us to examine the emission line profiles and kinematical morphologies in the central regions in more detail (positions 1 and 2 were taken in seeing of $\sim$0.5$''$). We present our results in the following.

\citet{monreal10a} and \citet{sidoli10} find broad line components in the supernebula region with widths of FWHM$\sim$80--150~\kms. In our analysis, we have been able to identify components with FWHM$>$250~\kms\ in a handful of spaxels. These very broad components are faint ($\sim$5 percent of the total flux); an example is shown in Fig.~\ref{fig:pos12_Ha_kinem}b. In these spaxels, the need for this additional component is very clear from the residuals and $\chi^2$ of the fit. In most of the field, however, the broad component line widths are consistent with those found in the aforementioned studies (FWHM$\sim$100--150~\kms, comprising 10--50 percent of the total line flux). 

\citet{monreal10a} identified a velocity gradient in the broad component extending over $\sim$$4\farcs7$ ($\sim$85~pc) centred on the supernebula and oriented NW-SE. This gradient is clearly seen in our C2 velocity map (Fig.~\ref{fig:pos12_Ha_kinem} bottom centre panel), and there are clear hints of this gradient also in the secondary narrow component (C3). Our IFU map shows that this gradient is present across the entire C2 map, with an appearance similar to a rotation field. However, there are distinct peaks on the red and blue side, hinting that there may be other explanations.

\begin{figure}
\centering
\includegraphics[width=0.5\textwidth]{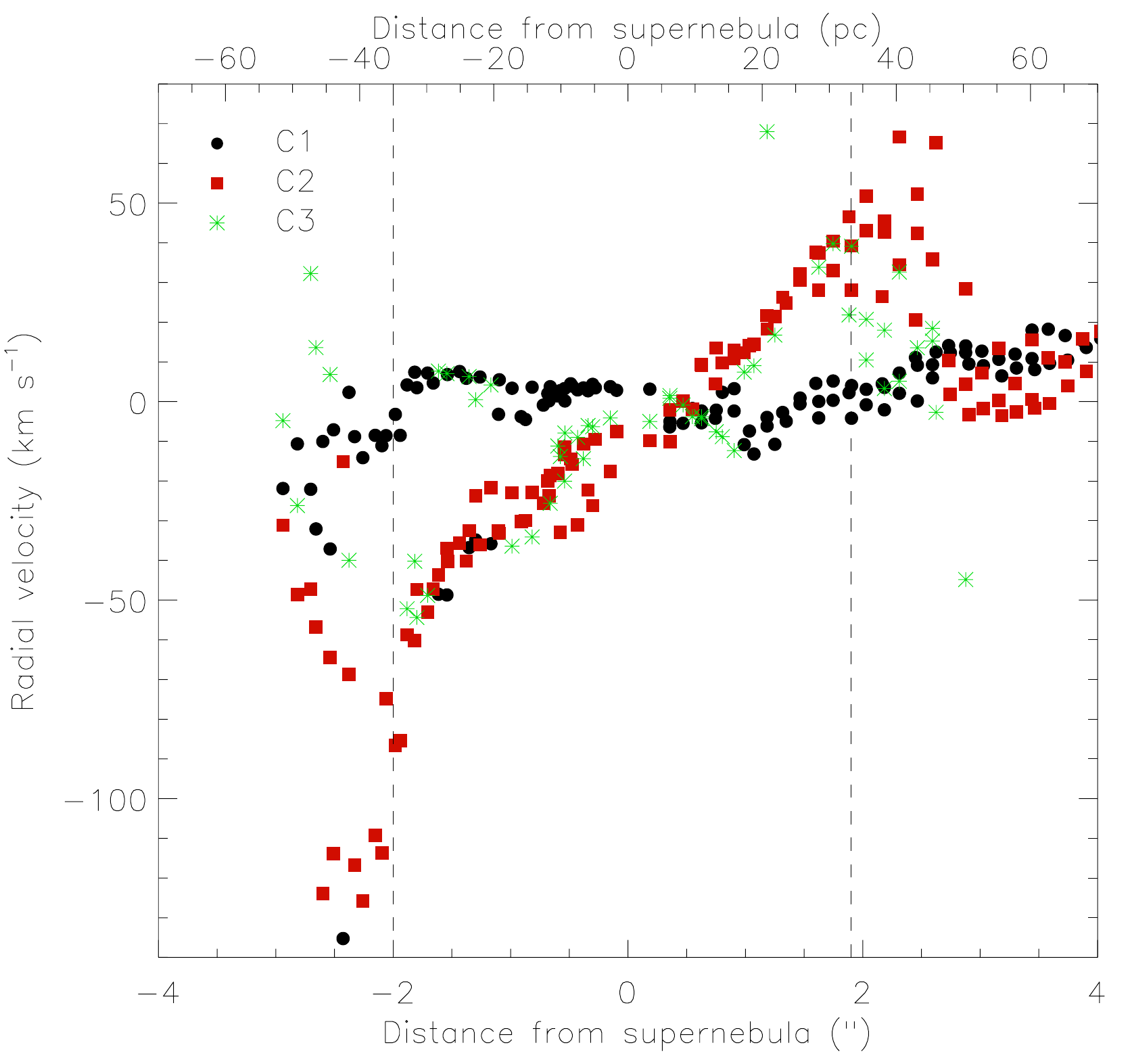}
\caption{Position-velocity diagram along the pseudoslit crossing the supernebula, and oriented along the maximum velocity gradient, as shown in Fig.~\ref{fig:pos3_Ha_kinem}. Velocities are shown relative to $v_{\rm sys}$ = $-390$~\kms\ \citep{schwartz04}. The vertical dashed lines indicate the region within which both C2 and C3 follow the same steep gradient. We interpret this as evidence of an outflow from the supernebula clusters.}
\label{fig:Ha_pv}
\end{figure}

To investigate this further, we defined a $0\farcs7$-wide pseudoslit crossing the supernebula, intersecting the two peaks, and oriented perpendicular to the line of nodes tracing zero velocity (i.e.\ along the maximum velocity gradient, as shown in Fig.~\ref{fig:pos12_Ha_kinem}). From this we extracted the radial velocities of the three line components to plot in Fig.~\ref{fig:Ha_pv}. From this figure, it is clear that whereas C1, the dominant narrow component, remains fairly static across the whole region with velocities varying by only $\sim$30~\kms, C2 and C3 trace a linear gradient in velocity from $\sim$$-50$ to $\sim$+50~\kms\ within the central $\sim$4$''$ (70~pc). Beyond this, C2 extends to even higher velocities, but is not followed by C3. All three components cross at zero velocity within $\sim$$0\farcs5$ of the location of the supernebula. \citet{r-r07} presented observations of the radio recombination line H53$\alpha$ over a similar region, and found a $\sim$50~\kms\ gradient over $\sim$$3\farcs5$ in the same NW-SE direction. That this is in very good agreement with our H$\alpha$ measurements indicates that the ionized material throughout the gas column in this region all behaves similarly.

\citet{monreal10a} interpreted this velocity gradient (seen by them only in the broad component) as tracing an outflow from the SSCs in the supernebula, and our data here also support this conclusion. The fact that the velocity gradient exists in the broad line component is difficult to reconcile with rotation since such a turbulent rotating structure could not maintain coherence for long. We therefore conclude that C2 and C3 trace an accelerating ionized gas outflow (although we cannot rule out the possibility that the outflow is superimposed on a more general rotation-like bulk motion existing separately to the majority of the ionized gas traced by component C1). The fact that the supernebula is located very close to the zero-point of the velocity gradient implies that the clusters embedded there are the driving sources of the outflow.

Proceeding with the outflow interpretation, its inclination, $i$, is still unknown; assuming $i$=45$^{\circ}$ the deprojected maximum outflow speed is $\sim$70~\kms. This velocity is roughly equivalent to the line width of C3 (FWHM$\sim$30--100~\kms), but is significantly less than the width of C2 (FWHM$\sim$100--250~\kms). This suggests that the turbulent energy in the outflow exceeds its bulk flow energy.

\subsubsection{Position 3}
IFU position 3 covers a region $\sim$6--10$''$ to the south-west of the supernebula containing two H\two\ knots, the eastern of which is associated with a known cluster of mass 2--3$\times$10$^{3}$~\Msol\ and age $<$4~Myr (H04-17). A number of other young star clusters not catalogued by \citet{harris04} are also located in the field.

C2 (the broad component) is identified across the entire field, and exhibits line widths extending to FWHM$>$200~\kms. The broadest line components are generally also blueshifted (example line profile (b) in Fig.~\ref{fig:pos3_Ha_kinem}). Around the western H$\alpha$ knot, the C2 component is fainter, narrower ($\sim$100~\kms), and redshifted compared to the surrounding material ($\sim$30~\kms); see example line profile (a). The corresponding C1 component is blueshifted, suggesting the gas is expanding or outflowing from this H$\alpha$ knot. To the south of the IFU field, we see regions of clear line splitting where the two components are more equally matched in terms of flux and width (example c).

This position partially overlaps with the IFU field-of-view of \citet{monreal10a}, and here our C1 radial velocities are in good agreement with their measurements.
They find similar regions around their knot 3 in the south-west of their field where two distinct narrow components are seen. They conclude that the environment here (clusters H04-3 and H04-5; see Fig.~\ref{fig:finder_WR}) must be more evolved than around the supernebula, and that these older clusters have cleared out their gas, leaving broken shells ionized by the remaining hot stars.

\subsubsection{Position 4}
Position 4 covers an interesting arc-shaped feature seen in H$\alpha$ $\sim$$12\farcs5$ (220~pc) to the south of the supernebula. There are a handful of young clusters in this area that appear to be located mostly along the rim of the H$\alpha$ arc, including one identified by \citet[][H04-27]{harris04}.

The very broad H$\alpha$ component seen in much of position 3 is not seen here. Line widths mostly remain below 100~\kms\ across the field, with only a few regions where line widths extend up to $\sim$150~\kms.  The H$\alpha$ line shape at the peak of emission is consistent with a single Gaussian. Across the rest of the arc, the line shape is characterised by a double-Gaussian (examples (b) and (c)) with velocity differences between the components of 20--30~\kms. In the north of the field, C2 is fainter and is blueshifted with respect to C1 by 20--30~\kms. These split line components indicate low-velocity flows of gas, likely to be shells or bubbles driven by the nearby star clusters. The bright H$\alpha$ knot located at the northern edge of the field does not appear distinct in the kinematics maps.

\subsubsection{Summary of kinematics}
Together with the results of \citet{monreal10a}, our results paint a picture of the overall gas dynamics in NGC 5253 consisting of localised gas flows, as part of multiple overlapping bubbles and filaments driven by the star clusters throughout the starburst. Close to the supernebula, we have been able to identify faint ($\sim$5 percent of the total flux) components with FWHM$>$250~\kms, although over most of positions 1 and 2, the broad component line widths are consistent with those found in the aforementioned studies \citep[FWHM$\sim$100--150~\kms, comprising 10--50 percent of the total line flux][]{sidoli10, monreal10a}. 

We confirm the presence of a strong velocity gradient in the central region seen by \citet{monreal10a}, extending over $\sim$$4\farcs5$ ($\sim$80~pc) oriented NW-SE direction. The p-v diagram extracted along the maximum gradient shows that both the broad and secondary narrow H$\alpha$ components (C2 and C3) both trace this steep gradient, which crosses zero velocity at the location of the supernebula. In line with \citet{monreal10a}, we interpret this as an ionized gas outflow from the supernebula clusters. Although the clusters are known to be deeply embedded \citep[$A_{V}$=7--17~mag;][]{vanzi04, alonsoherrero04}, this implies that the outflow has already carved channels in the dense gas distribution through which the ionized material is escaping. Furthermore, the fact that we see a continuous velocity gradient in both the blueshifted \emph{and} the redshifted side implies that the extinction in the supernebula must fall off rapidly such that both the front- and back-sides of the outflow are visible in the optical.

In our IFU position 3, 6--10$''$ to the south, the broad component contributes $<$15 percent, and by position 4 is not detected at all. This suggests that its presence is connected to the concentration of young clusters in the supernebula region, and most likely to the aforementioned outflow. 
However, since we detect broad emission $>$150~pc from this cluster it is unlikely to be solely due to its winds. It is possible that other young clusters are blowing winds that produce this broad line.  Since the broad lines in the central regions follow the bulk motions of the more quiescent gas, whatever causes this turbulence must be a local phenomenon. \citet{westm07a} detected a very similar broad line component to the ionized gas emission lines in NGC~1569. Here we found the broadest emission correlated with the location of gas clumps in the ISM, and hence concluded that it traces turbulent mixing layers \citep[TMLs;][]{slavin93, binette09} on the clump surfaces set up by the impact of stellar/cluster winds. In \citet{westm10b} we tested this hypothesis by observing a gas clump near to the Galactic OB association Pismis24 (NGC~6357) at high spatial and spectral resolution. That we found broad emission originating from the clump surface in projection confirmed the idea that TMLs can explain this kind of broad component emission.

\section{Chemical abundances} \label{sect:abund}

\begin{figure}
\centering
\includegraphics[width=0.3\textwidth]{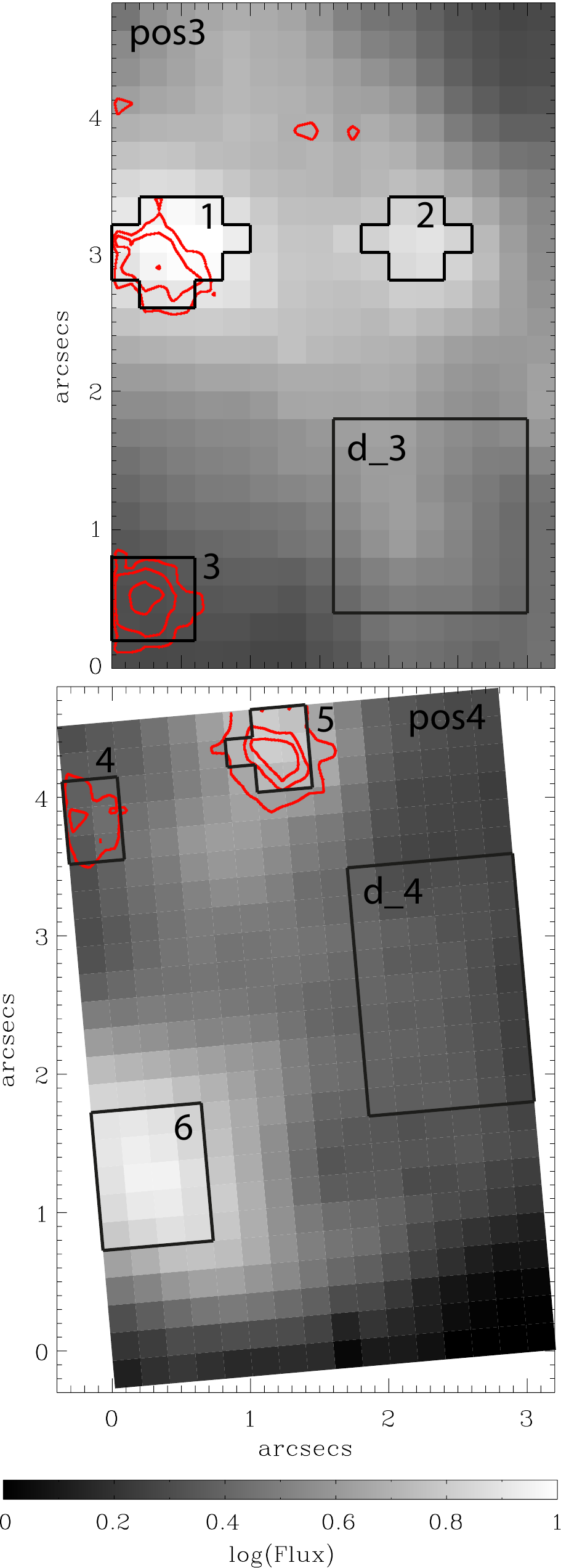}
\caption{IFU positions 3 and 4. Shown in greyscale is the integrated H$\alpha$ flux. Red contours show the integrated red WR bump flux (dominated by C\four $\lambda$5808 emission) originating from WC-type Wolf-Rayet stars. The black outlines show the 6 regions from which we extracted summed spectra for our abundance analysis.}
\label{fig:regions}
\end{figure}

Gas phase abundances in NGC~5253 have been of particular interest since the connection between N-enriched regions \citep{campbell86, walsh89, kobulnicky97a} and the presence of WR stars (WN-types in particular) was made \citep{schaerer97}. However, it is only recently that an unbiassed mapping of the abundances could be made using integral field spectroscopy \citep{monreal10a, monreal12}. Here, they studied a $\sim$9$\times$5~arcsec area encompassing the central regions. By mapping out the gas phase abundances at a resolution of $\sim$$0\farcs8$, they found two locations with enhanced N/O: one significant enhancement associated with the supernebula region and another much smaller enhancement associated with two clusters $\sim$7$''$ to the south-west (clusters H04-3 and H04-5).

\subsection{Definition of regions}
Our integral field observations provide the opportunity to investigate the gas phase abundances much further from the supernebula. However, the ionized gas surface brightness falls off rapidly with distance, and our observations are not of high enough S/N to derive spaxel-to-spaxel abundance maps as was done in \citet{monreal12}. Thus we have used the H$\alpha$ and WR flux maps to define precise apertures from which we extract integrated spectra and perform our analysis. This already represents a significant advantage over long-slit/fixed aperture studies, since positioning uncertainties and slit/aperture losses are avoided, and the measurements can be viewed in context since the surrounding environment is known. The location of our eight regions is shown in Fig.~\ref{fig:regions}. 

Region 1 ($\sim$$7\farcs4$ or 135~pc from the supernebula) corresponds to cluster H04-17 and to region He\two\ 4 from \citep{monreal10a}. The full spectrum from this region is shown in Fig.~\ref{fig:spectrum}. Region 2 corresponds to an H$\alpha$ knot $1\farcs8$ to the west; region 3 contains WR star signatures and likely corresponds to a cluster visible on the \textit{HST} image that was not catalogued by \citet[][see Figs.~\ref{fig:finder_WR} and \ref{fig:finder_WRpos12}]{harris04}; region 4 exhibits WR signatures, but no strong H$\alpha$ and there is no obvious cluster at this location in the \textit{HST} imaging; region 5 corresponds to an H$\alpha$ knot and most likely to a cluster not catalogued by \citet{harris04}; and region 6 corresponds to the H$\alpha$ knot associated with an arc or bubble. We also defined two larger boxes covering regions of fainter, diffuse gas, d\_3 and d\_4, in order to sample the abundances away from the bright H\two\ knots.

\subsection{Calculation of abundances}
Abundances were derived using \textsc{neat} \citep[Nebular Empirical Analysis Tool;][]{wesson12}\footnote{Full documentation, including source code and reference files, is available at \verb+http://www.sc.eso.org/~rwesson/codes/neat/+}, a code for calculating chemical abundances in photoionized nebulae. The code carries out a standard analysis of lists of emission lines using long-established techniques to estimate the level of interstellar extinction, calculate representative temperatures and densities, compute ionic abundances from both collisionally excited lines and recombination lines, and finally estimate total elemental abundances using an ionization correction scheme \citep{kingsburgh94}. Additionally, \textsc{neat} uses a Monte Carlo technique to robustly propagate uncertainties from line flux measurements and ionization correction factors (ICFs) through to the derived abundances. We note here that this more accurate treatment of the uncertainties produces reliable errors on the calculated abundances, and as such they may be larger than what can be found for equivalent measurements in the literature.

We measured the integrated line fluxes of all the major emission lines in each of the extracted integrated spectra and list them in Table~\ref{tab:fluxes}. We made no attempt to decompose each emission line into individual components for this part of the analysis. Using these values as inputs into \textsc{neat} we derived the electron temperature, \elt, the electron density, \eld, and ionic and elemental abundances for all six regions. The code was run with $\sim$10\,000 iterations, a number sufficient to establish a stable uncertainty distribution \citep{wesson12}. \elt's were derived from the temperature-sensitive \fnii$\lambda$5755/$\lambda$6584 line ratio, whilst \eld's were derived from the \fsii$\lambda$6716/$\lambda$6731 line ratio.  For Region~4, where \fnii$\lambda$5755 was undetected, we adopted the average \elt\ of Regions 5 and 6, 9700~K. Since the \foii$\lambda$3727 line was not within our wavelength coverage we adopted measurements from the nearest corresponding regions from \citet{walsh89}.  For our regions 1--3 we used the value I(\foii)/I(\hb)=3.083$\pm$0.034 \citep[corresponding to `region 2' of][]{walsh89}, whilst for our regions 4--6 we adopted I(\foii)/I(\hb)=2.929$\pm$0.031 \citep[`region 7' of][]{walsh89}. As mentioned above, \textsc{neat} adopts the ionisation correction factors (ICFs) listed by \citet{kingsburgh94} to convert from ionic to elemental abundances, and propagates the statistical uncertainties associated with the ICFs accordingly. Derived O, N, S, and He nebular abundances and their uncertainties are given in Table~\ref{tab:abundances}.

Our derived O abundances are, within the uncertainties, consistent with a metallicity of $\sim$0.2~\Zsol, as has been found in other regions of the galaxy (for a discussion see below). One point lies significantly above the others (for Region 5), but has large uncertainties that derive from the low S/N of the [N\two]$\lambda$5755 line in this region. The N abundances suggest a slightly lower metallicity. The N abundance in these outer regions is significantly lower than that found in the supernebula region, which is enhanced to $\sim$0.4~\Zsol\ \citep{monreal12}. Our He/H values are consistent with what was found by \citet{lopez-sanchez07}.

It is interesting to note how similar the spectra of the diffuse regions are to the emission knots/H\two\ regions. This implies that a significant fraction of the ionizing photons are leaking out of the gas clouds around the clusters to ionize the intra-[star]cluster medium.

\subsection{Radial abundance patterns}

\begin{figure}
\centering
\includegraphics[width=0.45\textwidth]{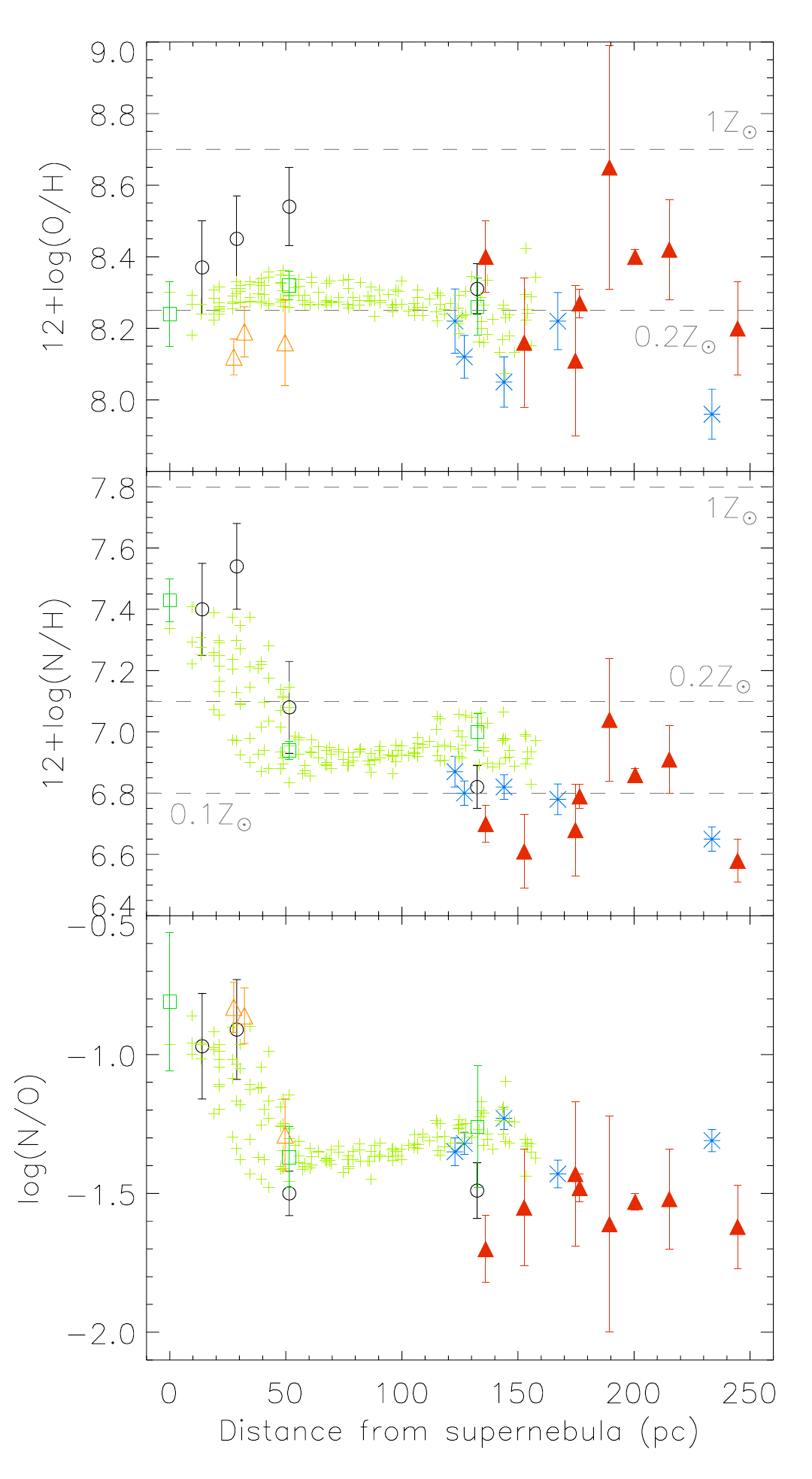}
\caption{O/H, N/H and log(N/O) abundances plotted against distance from supernebula. Plotted values correspond to those listed in Table~5; blue asterisks: \citet{walsh89}, orange triangles: \citet{kobulnicky97a}, black circles: \citet{lopez-sanchez07}, green squares: \citet{monreal12}, green plusses: \citet{monreal12} individual spaxels, red triangles: this work. Physical distances have been calculated assuming a distance of 3.8~Mpc \citep{sakai04}.}
\label{fig:abund_radial}
\end{figure}

We have compiled all the O/H, N/H and N/O abundance measurements for different regions found in the literature, and, together with our new measurements, listed them in Table~\ref{tab:radial_vals}. Fig.~\ref{fig:abund_radial} shows these values plotted versus distance from the supernebula (shown in Fig.~\ref{fig:finder} with a white cross). Note that one point at a particular radius is not necessarily associated with another point at the same/similar radius. Also included are the individual spaxel values from \citet[][their fig.~8]{monreal12}.  As mentioned above, (within the uncertainties) the O/H distribution is consistent with a metallicity of $\sim$0.2~\Zsol. Furthermore, we can see that it does not show any systematic trend (increasing or decreasing) within the 250~pc area sampled. The N/H profile, however, shows the well-known enhancement of $\sim$0.5~dex in log(N/H) within the central 50~pc. The large scatter of points within the rise result from the fact that the N-enhanced region is not circular in shape \citep{monreal12}. At larger radii, both our measurements and those of \citet{walsh89}\footnote{The uncertainties on the \citet{walsh89} measurements are small because they summed over much larger regions to get a high S/N and to explore possible similarities with more classical giant H\two\ regions.} are systematically lower than those measured by \citet{monreal12}, possibly due to the systematic uncertainties introduced by the different temperature- and density-sensitive lines used. However, given these uncertainties, the plot is consistent with a flat N abundance beyond the supernebular region (i.e.\ R$>$50~pc). The N/O distribution also reflects the N/H enhancement in the supernebula region, and is also consistent with being $\sim$flat beyond 50~pc.

Over the years a number of low-mass dwarf galaxies and giant H\two\ regions have been the subject of spectroscopic chemical abundance mapping \citep[e.g.][]{russell90, gonzalez-delgado94, kobulnicky96, devost97, kobulnicky97b, izotov99}, and more recently with IFUs \citep[e.g.][]{james09, james10, perez-montero11, lagos12, james12}. With very few exceptions (see below), these systems do not show significant internal chemical fluctuations or gradients, even though the expected yield of heavy elements produced by massive stars in young starbursts is substantial. \citet{kobulnicky97b} put forward two explanations: either ejecta from stellar winds and supernovae are transported throughout the galaxy and mixed instantaneously and uniformly, or freshly synthesised elements remain unmixed with the surrounding interstellar medium and reside in a hard-to-observe hot $10^6$~K phase (or a cold molecular phase). In support of the latter scenario, \citet{martin02} find evidence for a metal enriched diffuse X-ray phase in the halo of the starbursting dwarf galaxy NGC~1569. Measurements like this, however, are extremely hard to make with existing X-ray technology, and there are no equivalent X-ray measurements available for NGC~5253.

NGC~5253 is one of the most well-known (and certainly nearest) examples amongst the rare cases of dwarf galaxies with significant chemical variations. These include Mrk996 \citep{james09}, UM448 \citep{james12} and Haro11 \citep{james13}, possibly IC10 \citep{lopez-sanchez11} and IIZw70 \citep{kehrig08}, and those with radial oxygen abundance gradients \citep{lee06, werk10}. Of these, it is only NGC~5253 and Mrk996 that show a genuine N/H \textit{and} N/O enhancement (a high N/O ratio by itsself can result from the dilution of the O abundance by an inflow of metal-poor gas). However, although they both show these enrichments, the elevated N/H ratios in Mrk996 are only found in the broad emission-line component (FWHM$>$300~\kms), and the associated gas densities are even higher ($10^6$--$10^7$~\cmt). So what is it in NGC~5253 that puts it in this extremely rare class of galaxies?

\subsection{What is the origin of the N/O enhancement in the supernebula region?}
Putting together the available facts we know that:
\begin{itemize}
  \item The radio supernebula \citep{turner04} contains two very massive (combined mass $\sim$1--2$\times$$10^6$~\Msol), young (ages $\lesssim$4~Myr, but see below), embedded super star clusters (SSCs) \citep{vanzi04, alonsoherrero04, martin-hernandez05}.
  \item N/O is strongly enhanced in a region measuring $\sim$2$''$$\times4''$ extending in the NW-SE direction, and is not centred exactly on the supernebula cluster, but $\sim$0.5--1$''$ to the NW.
  \item There does not appear to be any obvious relation between the N-enhancements and the ages of the clusters and the WR populations. WR stars of both varieties are found up to 10$''$ from the SN, but there is no corresponding N enhancement at these locations. Apart from the supernebula region, only one other, much weaker, N/O enhancement was found around clusters H04-3 and 5 \citep{monreal12} with no coincident WR signatures.
  \item The density maps presented by \citet[][for [S\,{\sc ii]}]{monreal10a} and \citet[][for [O\,{\sc ii]}]{monreal12} show that the gas density peak (400--500~\cmt) is also elongated in the NW-SE direction, and centred to the NW of the supernebula.
  \item The gas pressure implied by these densities (assuming a temperature of $10^4$~K) is a few $\times10^6$~cm$^{-3}$~K. From radio observations, \citet{turner04} measure densities of 3--$4\times10^4$~\cmt, meaning that the pressure deep within the $<$1~pc size supernebula is certainly much higher.
  \item The H$\alpha$ line near the supernebula comprises a narrow/bright and broad/faint component, where the broad component has FWHM$\sim$100--280~\kms. The radial velocity map (Fig.~\ref{fig:pos12_Ha_kinem}) and pv diagram (Fig.~\ref{fig:Ha_pv}) show that there is a large velocity gradient along the same NW-SE direction traced by both the broad and secondary narrow H$\alpha$ components (C2 and C3). Our new data support the conclusion of \citet{monreal10a} that this is an ionized gas outflow, with the dynamical centre implying that the driving source(s) are the supernebula clusters (despite being so heavily embedded).
  \item \citet{beck12} observed [S\four]\,10.5\,\micron\ emission from the supernebula, and found a similar line profile to H$\alpha$, comprising a broad and narrow component. The broad component was $\sim$20~\kms\ blueshifted and interpreted as as indicating a gas outflow. The peak velocity of the [S\four] agrees with the Brackett lines \citep{turner03} and with H$\alpha$ (this work; although the H$\alpha$ broad component is broader and contributes less to the total line flux than [S\four]). Furthermore, spatially resolved observations of the radio-recombination line H53$\alpha$ over the same region show the same velocity gradient mentioned above \citep{r-r07}. These results suggest that H$\alpha$, Br$\alpha$, H53$\alpha$ and [S\four] are all tracing ionized material with approximately the same dynamics, despite originating at different depths within the nebula.
\end{itemize}

Since star clusters have been forming throughout the central regions of NGC~5253 for last $\sim$10~Myr \citep{harris04}, it is possible that previous generations have been responsible for the build-up in chemical abundances. However, the fact that the supernebula is located approximately in the centre of the chemically enriched region, and there is no evidence for older, sufficiently massive clusters in the vicinity strongly implies that the supernebula cluster(s) must be the source.

In line with above arguments regarding chemical fluctuations in dwarf galaxies, we propose that in NGC~5253, the chemically processed material expelled from star clusters in their early evolution remains in the hot wind phase and does not mix with cooler gas. However, the unusually high densities and pressures in the supernebula and surrounding medium have acted to stall or impede the winds from these embedded clusters, allowing time for the chemically enriched hot phase to mix with the cool/warm phases. Cluster winds stalled by high ambient pressures have been previously found in M82 \citep{smith06, silich07}.

Are the supernebula clusters old enough to be the enrichment source? The equivalent width of Pa$\alpha$ constrains the age of the clusters to $<$6~Myr \citep{alonsoherrero04, martin-hernandez05}. That the radio and millimetre continuum emission are almost entirely due to thermal emission \citep{beck96, turner00} suggests that no supernovae have exploded in the clusters, implying an upper age limit of $\sim$4~Myr. \citet{martin-hernandez05} modelled the mid-IR spectra of the embedded SSCs and found two solutions consistent with the observations: 1) a young ($<$4~Myr) cluster with a ``non-standard'', low, upper mass cutoff of the IMF; or 2) a cluster of $\sim$5--6~Myr with a ``standard'' high upper mass cutoff. The presence of N-enrichment strongly argues for the influence of WR stars, whose appearance in a cluster peaks between $\sim$3--5~Myr \citep{crowther07}. As \citet{martin-hernandez05} comments, one expects a supernova rate of at least 100 SN/Myr after approximately 3~Myr, but no effects of supernovae are seen in the radio. We therefore suggest that the clusters must be $\sim$4~Myr old -- old enough to have (had) a significant WR star population, but young enough to have had none or very few SNe.

That the thermal radio emission from the supernebula is optically thick classifies it as an ``ultra-dense'' H\two\ region with pressures significantly higher than typical interstellar values \citep{turner98}. Under normal circumstances this would imply that the clusters are extremely young ($\lesssim$1~Myr) from pressure/lifetime arguments \citep[cf.][]{kobulnicky99b, vacca02}. However, we now know that the clusters exist in a much larger high density/pressure environment, helping to explain why they remain so embedded even at ages of a few Myr \citep[cf.][]{smith06}.



The elongated morphology of the enriched region is spatially coincident with the extent and orientation of the outflow, implying that this has acted to increase the efficiency of either the mixing (to bring the new chemicals into the observable warm phase), and/or the subsequent diffusion. The outflowing N-enriched material has an extent of $\sim$60~pc, which could have propagated there over a few Myr with diffusion speeds of 20--60~\kms\ \citep{monreal10a}. This is in good agreement with the the outflow speeds measured above.

Further observations are required to explain why the densities/pressures in this supernebula region are so high. One possibility is that the diffuse H\one\ cloud infalling along the minor axis of the galaxy found by \citet{lopez-sanchez12} may be responsible for compressing the gas at this point.

%
%

It is interesting that the situation in the NGC~5253 supernebula is remarkably reminiscent of that in the only other system with a confirmed N/H enhancement: Mrk996 (D$\approx$22~Mpc). Here the N/H enhancement is seen in the nuclear region, spatially coincident with a significant WN star population \citep{james09}. Although the elevated N/H ratios are only found in a broad emission-line component (FWHM$>$300~\kms), the associated gas densities are also extremely high ($10^6$--$10^7$~\cmt).

\begin{table*}
\begin{center}
\begin{scriptsize}
\begin{threeparttable}
\begin{tabular}{l|r@{$\pm$}lr@{$\pm$}lr@{$\pm$}lr@{$\pm$}lr@{$\pm$}lr@{$\pm$}lr@{$\pm$}lr@{$\pm$}l}
\hline\hline																									
IFU position 3:	&	\multicolumn{4}{c}{Region 1}					&	\multicolumn{4}{c}{Region 2}					&	\multicolumn{4}{c}{Region 3}	&	\multicolumn{4}{c}{Diffuse Region}				\\
	&	\multicolumn{2}{c}{F$_{\rm obs}$}			&	\multicolumn{2}{c}{F$_{\rm dered}$}			&	\multicolumn{2}{c}{F$_{\rm obs}$}			&	\multicolumn{2}{c}{F$_{\rm dered}$}			&	\multicolumn{2}{c}{F$_{\rm obs}$}			&	\multicolumn{2}{c}{F$_{\rm dered}$} &	\multicolumn{2}{c}{F$_{\rm obs}$}			&	\multicolumn{2}{c}{F$_{\rm dered}$}			\\
\hline\hline																																	
\foii\, $\lambda$3727+29\tnote{a}	&	308.30	&	3.40	&	368.59	&	5.08	&	308.30	&	3.40	&	403.85	&	5.86	&	308.30	&	3.40	&	413.97	&	13.49	&	308.30	&	3.40	&	376.05	&	5.18	\\
\hb\,	&	100.00	&	0.89	&	100.00	&	0.89	&	100.00	&	0.82	&	100.00	&	0.82	&	100.00	&	3.10	&	100.00	&	3.10	&	100.00	&	0.93	&	100.00	&	0.93	\\
\foiii\, $\lambda$5007	&	523.74	&	3.32	&	510.84	&	3.29	&	361.04	&	2.11	&	347.68	&	2.08	&	373.89	&	8.24	&	358.81	&	8.06	&	336.34	&	2.20	&	327.14	&	2.18	\\
\ffeiii\, $\lambda$5270	&	0.63	&	0.03	&	0.59	&	0.03	&	0.88	&	0.12	&	0.79	&	0.10	&	1.81	&	0.18	&	1.61	&	0.16	&	0.87	&	0.04	&	0.81	&	0.04	\\
\fnii\, $\lambda$5755	&	0.30	&	0.03	&	0.26	&	0.03	&	0.51	&	0.12	&	0.42	&	0.09	&	0.77	&	0.11	&	0.62	&	0.09	&	0.53	&	0.02	&	0.46	&	0.02	\\
\hei\, $\lambda$5876	&	13.82	&	0.13	&	11.90	&	0.14	&	14.38	&	0.10	&	11.47	&	0.12	&	15.82	&	1.15	&	12.36	&	0.95	&	13.21	&	0.09	&	11.18	&	0.11	\\
\foi\, $\lambda$6300	&	4.69	&	0.04	&	3.88	&	0.05	&	6.32	&	0.09	&	4.76	&	0.09	&	10.91	&	0.41	&	8.00	&	0.39	&	8.05	&	0.06	&	6.52	&	0.08	\\
\fsiii\, $\lambda$6312	&	2.46	&	0.06	&	2.04	&	0.06	&	2.22	&	0.09	&	1.67	&	0.07	&	2.34	&	1.10	&	1.71	&	0.81	&	2.11	&	0.04	&	1.71	&	0.03	\\
\ha\,	&	380.36	&	2.58	&	303.62	&	3.77	&	399.67	&	2.33	&	284.30	&	3.77	&	413.20	&	9.12	&	284.88	&	12.69	&	365.38	&	2.39	&	284.38	&	3.50	\\
\fnii\ $\lambda$6584	&	19.89	&	0.49	&	15.67	&	0.42	&	29.88	&	0.29	&	20.83	&	0.33	&	33.17	&	1.14	&	22.38	&	1.20	&	31.11	&	0.22	&	23.86	&	0.31	\\
\hei\, $\lambda$6678	&	4.32	&	0.07	&	3.42	&	0.07	&	4.17	&	0.15	&	2.92	&	0.11	&	5.74	&	0.35	&	3.90	&	0.29	&	3.90	&	0.06	&	3.00	&	0.06	\\
\fsii\, $\lambda$6716	&	31.23	&	0.22	&	24.60	&	0.32	&	43.94	&	0.28	&	30.64	&	0.43	&	62.01	&	1.43	&	41.83	&	1.96	&	53.02	&	0.35	&	40.66	&	0.52	\\
\fsii\, $\lambda$6731	&	23.67	&	0.20	&	18.64	&	0.26	&	34.43	&	0.23	&	24.00	&	0.34	&	45.75	&	1.09	&	30.86	&	1.46	&	39.19	&	0.26	&	30.05	&	0.39	\\

c(\hb) &	\multicolumn{4}{c}{0.303$\pm$0.014}				&	\multicolumn{4}{c}{0.458$\pm$0.016}				&	\multicolumn{4}{c}{0.501$\pm$0.052}				&	\multicolumn{4}{c}{0.337$\pm$0.014}					\\

F(\hb) $\times 10^{-15}$ &	\multicolumn{4}{c}{15.75$\pm$0.20}  &	\multicolumn{4}{c}{4.67$\pm$0.010}	&	\multicolumn{4}{c}{1.45$\pm$0.1} &	\multicolumn{4}{c}{12.39$\pm$0.20}	\\
\hspace{0.5cm}erg~s$^{-1}$~cm$^{-2}$\\
\hline\hline

IFU position 4:	&	\multicolumn{4}{c}{Region 4}					&	\multicolumn{4}{c}{Region 5}					&	\multicolumn{4}{c}{Region 6}		&	\multicolumn{4}{c}{Diffuse Region}			\\
	&	\multicolumn{2}{c}{F$_{\rm obs}$}			&	\multicolumn{2}{c}{F$_{\rm dered}$}			&	\multicolumn{2}{c}{F$_{\rm obs}$}			&	\multicolumn{2}{c}{F$_{\rm dered}$}			&	\multicolumn{2}{c}{F$_{\rm obs}$}			&	\multicolumn{2}{c}{F$_{\rm dered}$} &	\multicolumn{2}{c}{F$_{\rm obs}$}			&	\multicolumn{2}{c}{F$_{\rm dered}$}			\\
\hline																																	
\foii\, $\lambda$3727+29\tnote{b}	&	292.90	&	3.10	&	352.87	&	10.08	&	292.90	&	3.10	&	335.43	&	5.03	&	292.90	&	3.10	&	336.22	&	4.77	&	292.90	&	3.10	&	357.27	&	5.98	\\
\hb\,	&	100.00	&	2.55	&	100.00	&	2.55	&	100.00	&	0.83	&	100.00	&	0.83	&	100.00	&	0.92	&	100.00	&	0.92	&	100.00	&	1.34	&	100.00	&	1.34	\\
\foiii\, $\lambda$5007	&	346.71	&	6.32	&	337.81	&	6.29	&	432.65	&	2.54	&	424.54	&	2.57	&	376.21	&	2.45	&	369.03	&	2.45	&	339.86	&	3.23	&	330.56	&	3.20	\\
\ffeiii\, $\lambda$5270	&	0.67	&	0.36	&	0.62	&	0.34	&	0.78	&	0.12	&	0.74	&	0.11	&	0.32	&	0.04	&	0.30	&	0.04	&	0.72	&	0.06	&	0.67	&	0.06	\\
\fnii\, $\lambda$5755	&	--	&	-- & --	&	--	&	0.28	&	0.12	&	0.25	&	0.11	&	0.31	&	0.04	&	0.28	&	0.04	&	0.37	&	0.06	&	0.32	&	0.05	\\
\hei\, $\lambda$5876	&	13.47	&	0.30	&	11.53	&	0.36	&	13.53	&	0.10	&	12.08	&	0.14	&	12.36	&	0.09	&	11.01	&	0.12	&	13.58	&	0.14	&	11.49	&	0.17	\\
\foi\, $\lambda$6300	&	9.69	&	0.32	&	7.96	&	0.35	&	5.68	&	0.09	&	4.92	&	0.10	&	2.31	&	0.04	&	2.00	&	0.04	&	6.82	&	0.08	&	5.53	&	0.10	\\
\fsiii\, $\lambda$6312	&	1.83	&	0.27	&	1.50	&	0.23	&	2.63	&	0.09	&	2.28	&	0.08	&	2.00	&	0.03	&	1.73	&	0.03	&	2.11	&	0.05	&	1.71	&	0.05	\\
\ha\,	&	362.26	&	6.60	&	286.39	&	10.91	&	344.89	&	2.03	&	290.66	&	4.25	&	337.15	&	2.19	&	283.30	&	3.84	&	378.06	&	3.59	&	294.24	&	5.57	\\
\fnii\ $\lambda$6584	&	29.01	&	0.89	&	22.62	&	1.06	&	21.77	&	0.27	&	18.16	&	0.34	&	16.92	&	0.14	&	14.07	&	0.21	&	29.60	&	0.31	&	22.70	&	0.46	\\
\hei\, $\lambda$6678	&	4.17	&	0.46	&	3.27	&	0.38	&	3.89	&	0.15	&	3.25	&	0.13	&	3.81	&	0.06	&	3.18	&	0.06	&	4.11	&	0.09	&	3.17	&	0.09	\\
\fsii\, $\lambda$6716	&	54.72	&	1.05	&	42.66	&	1.72	&	31.78	&	0.22	&	26.51	&	0.42	&	20.76	&	0.14	&	17.26	&	0.25	&	50.66	&	0.49	&	38.85	&	0.77	\\
\fsii\, $\lambda$6731	&	39.37	&	0.80	&	30.70	&	1.25	&	26.06	&	0.19	&	21.74	&	0.35	&	15.29	&	0.11	&	12.72	&	0.18	&	37.12	&	0.36	&	28.47	&	0.56	\\
																																	
c(\hb) &	\multicolumn{4}{c}{0.316$\pm$0.045}				&	\multicolumn{4}{c}{0.230$\pm$0.018}				&	\multicolumn{4}{c}{0.234$\pm$0.016}				&	\multicolumn{4}{c}{0.337$\pm$0.022}					\\
F(\hb) $\times 10^{-15}$ &	\multicolumn{4}{c}{1.31$\pm$0.10}	& \multicolumn{4}{c}{4.43$\pm$0.10}	&	\multicolumn{4}{c}{14.33$\pm$0.2} &	\multicolumn{4}{c}{10.75$\pm$0.5}	\\
\hspace{0.5cm}erg~s$^{-1}$~cm$^{-2}$\\
\hline
\end{tabular}
\begin{tablenotes}
\item[a]Values assumed from \citet{walsh89}, their region 2.
\item[b]Values assumed from \citet{walsh89}, their region 7.
\end{tablenotes}
\end{threeparttable}
\caption{Integrated emission line fluxes for the six defined regions within our IFU positions 3 and 4. Values are given relative to F(\hb)=100 (the \hb\ fluxes are given in the last rows), and de-reddenened using the c(\hb) values shown at the bottom. c(\hb) was derived from the \ha--\hb\ Balmer decrement, as described in the text.}
\label{tab:fluxes}
\end{scriptsize}
\end{center}
\end{table*}

\begin{table*}
\begin{center}
\begin{scriptsize}
\begin{threeparttable}
\title{NGC~5253}
\begin{tabular}{l|r@{$\pm$}lr@{$\pm$}lr@{$\pm$}lr@{$\pm$}lr@{$\pm$}lr@{$\pm$}lr@{$\pm$}lr@{$\pm$}l}
\hline
Quantity & \multicolumn{2}{c}{Region 1} & \multicolumn{2}{c}{Region 2} & \multicolumn{2}{c}{Region 3}& \multicolumn{2}{c}{Pos.3 Diffuse} & \multicolumn{2}{c}{Region 4} & \multicolumn{2}{c}{Region 5} & \multicolumn{2}{c}{Region 6} & \multicolumn{2}{c}{Pos.4 Diffuse} \\
\hline
N$_e$ &          96 &          17 & 
         153 &          17 & 
          58 &          42 & 
          56 &          12 & 
          35 &          30 & 
         208 &          48 & 
          57 &          20 & 
          51 &          20 \\
T$_e$(\fnii) &       10555 &         500 & 
       10564 &        1035 & 
       11775 &        1288 & 
       11187 &         219 & 
       9700 &           0 \tnote{a}& 
        8048 &         883 & 
       11378 &         758 & 
        9681 &         769 \\

        T$_e$(\foiii)\tnote{b} &       10194 &         906 & 
       11492 &        1639 & 
       11486 &        2102 & 
       10651 &         382 & 
       10000 &           0 & 
        7471 &        1323 & 
       10639 &         927 & 
        8850 &         815 \\
       [1.5ex]

O$^+/H^+\times10^4$ &    1.06 &    0.19 & 
    0.94 &    0.34 & 
    0.56 &    0.25 & 
    0.88 &    0.07 & 
    1.33 &    0.07 & 
    2.35 &    1.48 & 
    0.65 &    0.16 & 
    1.23 &    0.40 \\
O$^{++}/H^{+}\times10^4$ &    1.45 &    0.34 & 
    0.54 &    0.27 & 
    0.63 &    0.38 & 
    0.92 &    0.10 & 
    1.16 &    0.04 & 
    1.84 &    1.93 & 
    0.94 &    0.27 & 
    1.48 &    0.52 \\
O/H$\times10^4$ &    2.51 &    0.56 & 
    1.46 &    0.60 & 
    1.30 &    0.64 & 
    1.86 &    0.17 & 
    2.50 &    0.10 & 
    4.46 &    3.51 & 
    1.58 &    0.47 & 
    2.65 &    0.89 \\
    12+log(O/H) &    8.40 &    0.10 & 
    8.16 &    0.18 & 
    8.11 &    0.21 & 
    8.27 &    0.04 & 
    8.40 &    0.02 & 
    8.65 &    0.34 & 
    8.20 &    0.13 & 
    8.42 &    0.14 \\
           [1.5ex]

    N$^+/H^+\times10^6$ &    2.09 &    0.29 & 
    2.66 &    0.76 & 
    2.40 &    0.66 & 
    3.09 &    0.16 & 
    3.88 &    0.14 & 
    4.65 &    2.16 & 
    1.68 &    0.28 & 
    3.66 &    0.84 \\
N/H$\times10^6$ &    5.01 &    0.75 & 
    4.07 &    1.10 & 
    4.81 &    1.69 & 
    6.22 &    0.51 & 
    7.32 &    0.33 & 
   10.90 &    5.03 & 
    3.83 &    0.65 & 
    8.06 &    1.99 \\
    12+log(N/H) &    6.70 &    0.06 & 
    6.61 &    0.12 & 
    6.68 &    0.15 & 
    6.79 &    0.04 & 
    6.86 &    0.02 & 
    7.04 &    0.20 & 
    6.58 &    0.07 & 
    6.91 &    0.11 \\
    log(N/O) &   -1.70 &    0.12 & 
   -1.55 &    0.21 & 
   -1.43 &    0.26 & 
   -1.48 &    0.05 & 
   -1.53 &    0.03 & 
   -1.61 &    0.39 & 
   -1.62 &    0.15 & 
   -1.52 &    0.18 \\
          [1.5ex]

S$^+/H^+\times10^6$ &    0.75 &    0.08 & 
    0.94 &    0.23 & 
    1.03 &    0.26 & 
    1.21 &    0.06 & 
    1.62 &    0.04 & 
    1.75 &    0.61 & 
    0.47 &    0.07 & 
    1.37 &    0.25 \\
S$^{2+}/H^{+}\times10^6$ &    3.39 &    1.02 & 
    1.56 &    0.98 & 
    0.77 &    0.75 & 
    2.94 &    0.46 & 
    3.39 &    0.51 & 
    5.89 &    9.62 & 
    2.39 &    0.86 & 
    5.88 &    2.63 \\
S/H$\times10^6$ &    4.36 &    1.27 & 
    2.32 &    1.11 & 
    1.89 &    1.07 & 
    4.34 &    0.55 & 
    5.23 &    0.54 & 
    8.48 &   11.30 & 
    3.03 &    0.99 & 
    6.95 &    2.86 \\
12+log(S/H) &    6.64 &    0.13 & 
    6.37 &    0.21 & 
    6.28 &    0.24 & 
    6.64 &    0.05 & 
    6.72 &    0.04 & 
    6.93 &    0.57 & 
    6.48 &    0.14 & 
    6.84 &    0.18 \\
log(S/O) &   -1.76 &    0.16 & 
   -1.80 &    0.27 & 
   -1.84 &    0.32 & 
   -1.63 &    0.07 & 
   -1.68 &    0.05 & 
   -1.72 &    0.67 & 
   -1.72 &    0.19 & 
   -1.58 &    0.23 \\
          [1.5ex]

He/H$\times10^2$ &    8.13 &    0.09 & 
    7.99 &    0.05 & 
    8.99 &    0.55 & 
    7.91 &    0.07 & 
    8.19 &    0.29 & 
    8.19 &    0.29 & 
    7.92 &    0.07 & 
    7.92 &    0.11 \\
\hline
W$_{\rm eq}$(H$\beta$) (\AA) &    $-96.55$&   5 & 
    $-60.12$&   4 & 
    $-14.06$&   2 & 
    $-55.43$&   3 & 
    $-33.52$&   5 & 
    $-105.35$&   10 & 
$-99.88$ & 10 &
$-42.41$ & 5 \\
L(WR bump) & 10.65 & 1.17 &
    \multicolumn{2}{c}{} &
    9.05 & 1.17 &
    \multicolumn{2}{c}{}&
    4.91 & 1.17 &
    8.49 & 1.17 &
 \multicolumn{2}{c}{} &
  \multicolumn{2}{c}{} \\
\hspace{0.1cm} ($\times$10$^{35}$ erg~s$^{-1}$) \\
Equivalent no.\ of & \multicolumn{2}{c}{$\sim$1} &
    \multicolumn{2}{c}{} &
    \multicolumn{2}{c}{$\sim$1} &
     \multicolumn{2}{c}{} &
    \multicolumn{2}{c}{$\sim$1} &
    \multicolumn{2}{c}{$\sim$1} &   
\multicolumn{2}{c}{}& 
\multicolumn{2}{c}{}\\
\hspace{0.1cm} WCE\mtnote{c}~stars \\

\hline
\end{tabular}
\begin{tablenotes}
\item[a] \fnii$\lambda$5755 was undetected in this region, hence we adopted the average \elt\, from the surrounding regions of $\sim$9700~K.
\item[b] Derived from Perez-Montero relationship.  See text for details.
\item[c] Early-type WC stars (WCE) are expected to dominate at the metallicity of NGC~5253. To estimate the number of WC stars in each region, we used the relation given by \citet{lopez-sanchez10} for the C\four\,$\lambda$5808 luminosity of a WCE star [i.e.\ L(WR bump)] at the metallicity of NGC~5253. The luminosities we measure are consistent with the presence of 1 WCE star in each source.

\end{tablenotes}
\end{threeparttable}
\caption{Abundance measurements for the six defined regions within IFU positions 3 and 4. Also given are the equivalent widths of H$\beta$ (W$_{\rm eq}$(H$\beta$), and the integrated red WR bump luminosities that provide an estimate of the equivalent number of WCE stars.}
\label{tab:abundances}
\end{scriptsize}
\end{center}
\end{table*}

\begin{table}
\label{tab:radial_vals}
\caption{O/H, N/H and N/O abundance ratios from the literature and this work for various regions across the starburst. The nomenclature is from the respective papers. These values were used to create the plots in Fig.~\ref{fig:abund_radial}.}
\centering
\scriptsize
\begin{threeparttable}
\begin{tabular}{lcccc}
\hline
	&	D\mtnote{a}  &	12+log(O/H)			&	12+log(N/H)		&	log(N/O)			\\
	&	 arcsec(pc)	\\
	\hline
\multicolumn{5}{l}{Walsh \& Roy (1989)}														\\
Region 2	&	6.68(123)	&	8.22	$\pm$	0.09	&	6.87	$\pm$	0.05	&	$-1.35$	$\pm$	0.05	\\
Region 4	&	9.09(167)	&	8.22	$\pm$	0.08	&	6.78	$\pm$	0.05	&	$-1.43$	$\pm$	0.05	\\
Region 5	&	6.90(127)	&	8.12	$\pm$	0.06	&	6.80	$\pm$	0.04	&	$-1.32$	$\pm$	0.04	\\
Region 6	&	7.83(144)	&	8.05	$\pm$	0.07	&	6.82	$\pm$	0.04	&	$-1.23$	$\pm$	0.04	\\
Region 7	&	12.70(234	)&	7.96	$\pm$	0.07	&	6.65	$\pm$	0.04	&	$-1.31$	$\pm$	0.04	\\[1ex]
\multicolumn{5}{l}{Kobulnicky et al.\ (1997)}		\\
HII-1		&	1.50(28)	&	8.12	$\pm$	0.05	&	---\mtnote{b}		&	$-0.83$	$\pm$	0.09 \\
HII-2		&	1.75(32)	&	8.19	$\pm$	0.07	&	---\mtnote{b}		&	$-0.86$	$\pm$	0.10 \\
UV1    	&	2.70(50)  	&	8.16	$\pm$	0.12	&	---\mtnote{b}		&	$-1.29$	$\pm$	0.13 \\[1ex]
\multicolumn{5}{l}{Lopez-Sanchez et al.\ (2007)}														\\
Knot A	&	1.57(29)	&	8.45	$\pm$	0.12	&	7.54	$\pm$	0.14	&	$-0.91$	$\pm$	0.18	\\
Knot B	&	0.76(14)	&	8.37	$\pm$	0.13	&	7.40	$\pm$	0.15	&	$-0.97$	$\pm$	0.19	\\
Knot C	&	2.80(52)	&	8.54	$\pm$	0.11	&	7.08	$\pm$	0.15	&	$-1.50$	$\pm$	0.08	\\
Knot D	&	7.20(132)	&	8.31	$\pm$	0.07	&	6.82	$\pm$	0.07	&	$-1.49$	$\pm$	0.10	\\ [1ex]
\multicolumn{5}{l}{Monreal-Ibero et al.\ (2012)}														\\
Knot 1	&	0.00(0)	&	8.24	$\pm$	0.09	&	7.43	$\pm$	0.07	&	$-0.81$	$\pm$	0.25	\\
Knot 2	&	2.80(51)	&	8.32	$\pm$	0.04	&	6.94	$\pm$	0.03	&	$-1.37$	$\pm$	0.11	\\
Knot 3	&	7.21(133)	&	8.26	$\pm$	0.08	&	7.00	$\pm$	0.06	&	$-1.26$	$\pm$	0.22	\\[1ex]
\multicolumn{5}{l}{This work - Position 3}														\\
Region 1	&	7.4(136)	&	8.40	$\pm$	0.10	&	6.70	$\pm$	0.06	&	-1.70	$\pm$	0.12	\\
Region 2	&	8.3(153)	&	8.16	$\pm$	0.18	&	6.61	$\pm$	0.12	&	-1.55	$\pm$	0.21	\\
Region 3	&	9.5(175)	&	8.11	$\pm$	0.21	&	6.68	$\pm$	0.15	&	-1.43	$\pm$	0.26	\\
Region d\_3	&	9.6(177)	&	8.27	$\pm$	0.04	&	6.79	$\pm$	0.04	&	-1.48	$\pm$	0.05 \\[1ex]
\multicolumn{5}{l}{This work - Position 4}														\\
Region 4	&	10.9(200)	&	8.40	$\pm$	0.02	&	6.86	$\pm$	0.02	&	-1.53	$\pm$	0.03	\\
Region 5	&	10.3(189)	&	8.65	$\pm$	0.34	&	7.04	$\pm$	0.20	&	-1.61$\pm$	0.39	\\
Region 6	&	13.3(245)	&	8.20	$\pm$	0.13	&	6.58	$\pm$	0.07	&	-1.62	$\pm$	0.15	\\
Region d\_4	&	11.7(215)	&	8.42	$\pm$	0.14	&	6.91	$\pm$	0.11	&	-1.52	$\pm$	0.18 \\
\hline
\end{tabular}
\begin{tablenotes}
\item[a] Projected distance from the supernebula.
\item[b] These authors measured the N/O ratio directly from the intensity of [N\two]$\lambda$6584 and [O\two]$\lambda$3727 since they did not detect sufficient lines to measure the N abundance itself.
\end{tablenotes}
\end{threeparttable}
\normalsize
\end{table}

\section{Conclusions and Summary} \label{sect:summary}

We obtained Gemini-S/GMOS-IFU observations of four regions near the centre of the nearby dwarf starburst galaxy NGC 5253. We identify 11 distinct WCE-type stars through mapping out the strength and location of the red C\four\ WR bump. Some are associated with known young clusters catalogued by \citet{harris04}, and two are coincident with the location of WN-type WR stars detected by \citet{monreal10a}. The finding of WR stars spread out over 20$''$ ($\sim$350~pc) attests to the large area over which the starburst has occurred.

We fitted multiple Gaussian profile models to the emission lines to recover kinematic information about the ionized gas in the four IFU positions. Together with the results of \citet{monreal10a}, our results paint a picture of the gas dynamics in NGC~5253 consisting of localised gas flows, as part of multiple overlapping bubbles and filaments driven by the star clusters throughout the starburst. We identify a broad (FWHM$>$100~\kms) line component; near the famous supernebula \citep{turner03} it contributes 10--50 percent to the total line flux, but this falls off with distance, and is not detected at all in IFU position 4 ($>$10$''$ to the south). This suggests that its origin is connected to the concentration of young clusters in the supernebula region.

We confirm the presence of a strong velocity gradient over $\sim$$4\farcs5$ ($\sim$80~pc), centred on the location of the supernebula, as seen by \citet{monreal10a}, and show that both the broad and secondary narrow H$\alpha$ components follow the same pattern. In line with \citet{monreal10a}, we interpret this as an accelerating ionized gas outflow from the supernebula clusters.

We extracted integrated spectra and measured ionized-phase abundances of O, N and He in a number of regions at distances $7\farcs4$--$13\farcs3$ from the supernebula. By combining these with O/H, N/H and N/O measurements of nearer regions found in the literature, we found that, within the central 250~pc, the O/H and N/H distributions are flat and consistent with a metallicity of $\sim$0.2~\Zsol. It is only within the central 50~pc that the nitrogen abundance is strongly enhanced by log(N/H)$\sim$0.5~dex.

This enhancement is what puts NGC~5253 in the rare class of systems with significant chemical variations. That the famous supernebula is located approximately in the centre of the N-enriched region, and there is no evidence for older, sufficiently massive clusters in the vicinity, strongly implies that the supernebula cluster(s) must be the source of the extra N. The presence of this enrichment strongly argues for the influence of WR stars and thus an age for the supernebula clusters of $\sim$3--5~Myr; the lack of evidence for supernovae argues for ages $\lesssim$4~Myr  \citep{beck96, turner00}. 

We propose that the unusually high densities and pressures in the supernebula and surrounding medium have acted to impede the winds from the supernebula clusters, allowing time for the chemically enriched hot wind material to mix with the cool/warm phases (and therefore become visible in the optical/near-IR). Under normal conditions, this enriched wind material remains in the hot wind phase and does not mix with cooler gas \citep[as described by][]{kobulnicky97b}, thus explaining why the rest of the galaxy (and most other dwarf galaxies) is chemically uniform in the warm ionized gas phase. We argue that the outflow has acted to spread the injected N-rich gas from the WR winds and helped to increase the efficiency of the hot-cool phase mixing to allow it to be seen in the optical.

\section*{Acknowledgments}
We acknowledge the anonymous referee for their useful comments. 
The research leading to these results has received funding from the European Community's Seventh Framework Programme (/FP7/2007-2013/) under grant agreement No 229517. AM-I is supported by the Spanish Research Council within the program JAE-Doc, Junta para la Amplicaci\'{o}n de Estudios, co-funded by the FSE and by the projects AYA2010-21887 from the Spanish PNAYA, CSD2006 - 00070 ``1st Science with GTC'' from the CONSOLIDER 2010 programme of the Spanish MICINN, and TIC114 Galaxias y Cosmolog\'{i}a of the Junta de Andaluc\'{i}a (Spain). AM-I would also like to acknowledge support from the ESO Visitor Programme. MSW would like to thank Nate Bastian for discussions regarding IMF sampling and Paul Crowther regarding WR stars.

\footnotesize{
\bibliographystyle{aa}
\bibliography{/Users/mwestmoq/Dropbox/Work/references}
}

\label{lastpage}
\end{document}